%% file: SUS-21-008_temp.tex
\pdfoutput=1
\documentclass[11pt,twoside,a4paper,cmspaper,final,collab]{cms-tdr}
\def\svnVersion{c11bad6}\def\svnDate{2024/06/06}\def\cmsCernNoTag{CERN-EP-2023-238}\def\cmsCernDate{\today}\def\cmsMessage{Published in Physical Review D as \href{http://dx.doi.org/10.1103/PhysRevD.109.112001}{\doi{10.1103/PhysRevD.109.112001}.}}
\begin{document}\cmsNoteHeader{SUS-21-008}

\newlength\cmsTabSkip\setlength{\cmsTabSkip}{1ex}

\newcommand{\multilep}{``2SS\ensuremath{\Pell/{\geq}\,3\Pell}"\xspace}
\newcommand{\multilepNoQuotes}{2SS\ensuremath{\Pell/{\geq}\,3\Pell}\xspace}
\newcommand{\slep}{{\HepSusyParticle{\Pell}{}{}}\Xspace}
\newcommand{\nino}{\PSGcz}
\newcommand{\ninoone}{\PSGczDo}
\newcommand{\ninotwo}{\PSGczDt}
\newcommand{\ninothree}{\PSGczDT}
\newcommand{\chinoone}{\PSGcpmDo}
\newcommand{\chinoonemp}{\ensuremath{\PSGc_1^\mp}\xspace}
\newcommand{\mninoone}{\ensuremath{m_{\PSGczDo}}\xspace}
\newcommand{\mninotwo}{\ensuremath{m_{\PSGczDt}}\xspace}
\newcommand{\mchinoone}{\ensuremath{m_{\PSGcpmDo}}\xspace}
\newcommand{\lumiFull}{137\fbinv}
\newcommand{\gravitino}{\PXXSG}
\newcommand{\ww}{\ensuremath{\PW\PW}\xspace}
\newcommand{\wz}{\ensuremath{\PW\PZ}\xspace}
\newcommand{\wx}{\ensuremath{\PW\mathrm{X}}\xspace}
\newcommand{\wh}{\ensuremath{\PW\PH}\xspace}
\newcommand{\zz}{\ensuremath{\PZ\PZ}\xspace}
\newcommand{\zh}{\ensuremath{\PH\PZ}\xspace}
\newcommand{\hh}{\ensuremath{\PH\PH}\xspace}
\newcommand{\pp}{\ensuremath{{\Pp\Pp}}\xspace}
\newcommand{\mll}{\ensuremath{m_{\Pell\Pell}}\xspace}
\newcommand{\mt}{\ensuremath{M_\text{T}}\xspace}
\newcommand{\dm}{\ensuremath{\Delta m}\xspace}
\newcommand{\ptmisscorr}{\ensuremath{\pt^\text{miss, corr}}\xspace}
\newcommand{\mtlll}{\ensuremath{M_{\text{T}}^{3\Pell}}\xspace}
\newcommand{\sutwo}{\ensuremath{\text{SU}(2)_{\text{L}}}\xspace}
\newcommand{\drmax}{\ensuremath{\Delta R_{\text{max}}}\xspace}
\ifthenelse{\boolean{cms@external}}{\providecommand{\cmsLeft}{upper\xspace}}{\providecommand{\cmsLeft}{left\xspace}}
\ifthenelse{\boolean{cms@external}}{\providecommand{\cmsRight}{lower\xspace}}{\providecommand{\cmsRight}{right\xspace}}
\ifthenelse{\boolean{cms@external}}{\providecommand{\cmsLEFT}{Upper\xspace}}{\providecommand{\cmsLEFT}{Left\xspace}}
\ifthenelse{\boolean{cms@external}}{\providecommand{\cmsRIGHT}{Lower\xspace}}{\providecommand{\cmsRIGHT}{Right\xspace}}
\ifthenelse{\boolean{cms@external}}{\providecommand{\cmsTable}[1]{#1}}{\providecommand{\cmsTable}[1]{\resizebox{0.9\textwidth}{!}{#1}}}

\cmsNoteHeader{SUS-21-008}
\title{Combined search for electroweak production of winos, binos, higgsinos, and sleptons in proton-proton collisions at \texorpdfstring{$\sqrt{s} = 13\TeV$}{sqrt(s) = 13 TeV}}

\date{\today}

\abstract{A combination of the results of several searches for the electroweak production of the supersymmetric partners of standard model bosons, and of charged leptons, is presented. All searches use proton-proton collision data at $\sqrt{s}=13\TeV$ recorded with the CMS detector at the LHC in 2016--2018. The analyzed data correspond to an integrated luminosity of up to 137\fbinv. The results are interpreted in terms of simplified models of supersymmetry. Two new interpretations are added with this combination:  a model spectrum with the bino as the lightest supersymmetric particle together with mass-degenerate higgsinos decaying to the bino and a standard model boson, and the compressed-spectrum region of a previously studied model of slepton pair production. Improved analysis techniques are employed to optimize sensitivity for the compressed spectra in the wino and slepton pair production models. The results are consistent with expectations from the standard model. The combination provides a more comprehensive coverage of the model parameter space than the individual searches, extending the exclusion by up to 125\GeV, and also targets some of the intermediate gaps in the mass coverage.}

\hypersetup{
pdfauthor={CMS Collaboration},
pdftitle={Combined search for electroweak production of winos, binos, higgsinos, and sleptons in proton-proton collisions at sqrt(s) = 13 TeV},
pdfsubject={CMS},
pdfkeywords={CMS, SUSY, winos, binos, higgsinos, sleptons, electroweak production}}

\maketitle

\section{Introduction}
The standard model (SM), despite its success in describing the fundamental particles and their interactions, leaves several open questions in particle physics unanswered. Various extensions of the SM, such as supersymmetry (SUSY)~\cite{SUSY0,Nilles:1983ge,SUSY2,SUSY3,SUSY4}, have been developed to address these questions. Supersymmetry potentially provides a solution to the hierarchy problem~\cite{Dimopoulos:1995mi,Barbieri:2009ev,Papucci:2011wy}, as well as the unification of the gauge couplings at high energy scales~\cite{unification,SUSY2}. Moreover, if $R$-parity~\cite{Farrar:1978xj} is conserved, the lightest SUSY particle (LSP) would be stable, and a potential dark matter candidate~\cite{Goldberg:1983nd,Ellis:1983ew}.

Supersymmetry introduces a fermi\-onic (bos\-onic) superpartner for each boson (fermion) of the SM. 
The superpartners of the leptons are called sleptons; $\slep$ denotes the superpartner of lepton \Pell. Those of the SM \sutwo and U(1) gauge fields before electroweak (EW) symmetry breaking are the winos and binos, respectively, collectively called gauginos.
In the minimal SUSY theory, MSSM~\cite{Fayet:1976cr,Nilles:1983ge,Martin:1997ns}, a new complex Higgs doublet is added to the SM.  The MSSM thus contains five Higgs bosons along with the four higgsino superpartners of the two Higgs doublets.  
The bino, winos, and higgsinos can mix among one another to form in total eight mass eigenstates (collectively, electroweakinos): two chargino pairs ($\PSGc_{1,2}^\pm$) and four neutralinos ($\PSGc_{\text{1-4}}^0$). 
While in some models \ninoone is taken as the LSP, this assumption is not required.
For example, models motivated by gauge-mediated SUSY breaking (GMSB)~\cite{Dimopoulos:1996vz,Matchev:1999ft} introduce a Nambu--Goldstone boson goldstino (\sGra)
that may be identified with two of the chirality components of the superpartner of a graviton (gravitino), and be considered the LSP.

Interactions among the charginos, neutralinos, and sleptons occur with the same EW couplings as those of their SM partners.  At the CERN LHC, the cross sections for production of these particles are correspondingly small compared with those for the SUSY partners of the strongly interacting SM particles (squarks and gluinos).  Nevertheless, if the squarks and gluinos are more massive than the EW SUSY particles, the EW superpartners might be the only SUSY particles accessible at the LHC.

The ATLAS~\cite{Aad:2012jva,Aad:2015jqa,ATLAS:2015eiz,Aaboud:2017dmy,Aaboud:2017hdf,ATLAS:2017qwn,Aaboud:2017leg,Aaboud:2018doq,ATLAS:2018rns,Aaboud:2018jiw,Aaboud:2018sua,Aaboud:2018htj,Aaboud:2018ngk,Aad:2019qnd,Aad:2019vvi,Aad:2020qnn,Aad:2019vvf,ATLAS:2019gti,Aad:2019qnd,Aad:2019vnb,ATLAS:2021yyr,ATLAS:2021moa,ATLAS:2021yqv,ATLAS:2022zwa,ATLAS:2022hbt,ATLAS:2023lfr,ATLAS:2023act,ATLAS:2024tqe,ATLAS:2024qxh} and CMS~\cite{Khachatryan:2014qwa,Chatrchyan:2013mya,Khachatryan:2014mma,Sirunyan:2017uyt,Sirunyan:2017zss,Sirunyan:2017yse,Sirunyan:2017eie,Sirunyan:2017obz,Sirunyan:2018ubx,Sirunyan:2017qaj,Sirunyan:2017qaj,Sirunyan:2017mrs,Sirunyan:2017hvp,Sirunyan:2017nyt,Sirunyan:2017bsh, CMS:2017moi, Sirunyan:2018ubx,Sirunyan:2018iwl,Sirunyan:2018lul,Sirunyan:2018psa,Sirunyan:2019hzr,CMS:2019san,Sirunyan:2019mbp,Sirunyan:2019sem,Sirunyan:2019iwo,Sirunyan:2019xwh,CMS:2019zmn,CMS:2019eln,Sirunyan:2020eab,Sirunyan:2020eab,CMS:2021few,CMS:2021edw,CMS:2021cox, CMS:2022vpy, CMS:2022sfi, CMS:2023xlp} Collaborations have carried out extensive search programs that target final states that could result from the production and decay of EW interacting SUSY particles.  The proton-proton (\pp) collision energy $\sqrt{s}$ for Ref.~\cite{Aad:2012jva} (ATLAS) was 7\TeV; for Refs.~\cite{Aad:2015jqa,ATLAS:2015eiz} (ATLAS) and~\cite{Khachatryan:2014qwa,Chatrchyan:2013mya,Khachatryan:2014mma} (CMS) it was 8\TeV; for all other searches it was 13\TeV. 
Given that the SUSY particle decays could result in more than one final state, these programs benefit from combining individual searches to maximally exploit the available data, and both collaborations have performed combined searches for EW particles at $\sqrt{s}=8\TeV$, by ATLAS~\cite{ATLAS:2015eiz} and CMS~\cite{Khachatryan:2014qwa,Khachatryan:2014mma}, and at $13\TeV$, by ATLAS~\cite{ATLAS:2021moa,ATLAS:2024qxh} and CMS~\cite{Sirunyan:2018ubx}.

In this paper we present the results of a combination of the EW SUSY searches reported in Refs.~\cite{Sirunyan:2020eab,CMS:2021few, CMS:2021edw, CMS:2021cox, CMS:2022vpy, CMS:2022sfi}. 
The data were collected at the LHC in 2016--2018 and correspond to an integrated luminosity of \lumiFull, except as noted in the following sections. 
The interpretation of results was performed using simplified model spectra (SMS) of SUSY~\cite{bib-sms-1,bib-sms-2,bib-sms-3,bib-sms-4}, which assume that all SUSY particles other than those directly involved in the specified process are decoupled, being too massive to be produced or be involved in the decays. 
Additionally, $R$-parity is assumed to be conserved, ensuring that the initial production process gives rise to pairs of superpartners. 
Both compressed and uncompressed mass spectra scenarios were targeted; in the former the mass-splitting \dm between the lightest and next-to-lightest supersymmetric particles is $\mathcal{O}(10)\GeV$ and smaller, resulting in an experimentally challenging final state in which the 
observable decay products have low momentum.
Here we present an improved and extended reanalysis of the search of Ref.~\cite{CMS:2021edw}: a novel signal extraction method is employed, along with a signal selection optimized to search for slepton pair production in models with compressed spectra. 
We also consider two additional interpretations beyond those covered in the Refs.~\cite{Sirunyan:2020eab,CMS:2021few,CMS:2021edw,CMS:2021cox, CMS:2022vpy, CMS:2022sfi}, one of them enabled by the updates provided for the search of Ref.~\cite{CMS:2021edw}. 
Together with the increased sensitivity resulting from the combined search itself, these provide a more comprehensive coverage of the model parameter space than the original individual searches.  Tabulated results are provided in the HEPData record for this analysis~\cite{hepdata}.

This paper is organized as follows. The specific SMS scenarios studied in this combination are discussed in Section~\ref{sec:signalModels}. The CMS detector is briefly described in Section~\ref{sec:detector}, while Section~\ref{sec:EventRecoMC} contains summaries of 
the event reconstruction methods as well as the simulation of the different background and signal processes. 
Individual searches used as input are detailed in Section~\ref{sec:searches}, together with the updates provided for this combination where applicable. 
The general strategy used to combine the input analyses is described in Section~\ref{sec:combStrat}, followed by the description of the treatment of systematic uncertainties in Section~\ref{sec:syst}. 
The interpretation of the combined data under the considered SMS scenarios is provided in Section~\ref{sec:interpretations}, and a summary of the results is provided in Section~\ref{sec:summary}.

\section{Signal models and search strategy}
\label{sec:signalModels}

In this section we introduce the specific SUSY models used to interpret the results of the combined search, together with a summary of the component searches.
Each search is characterized by its topology, that is, the combination of SM particles directly emitted in the decay chains of SUSY particles in a given production and decay process. For example, the associated production of \chinoone and \ninotwo, with the \chinoone (\ninotwo) decaying to a \PW (\PZ) boson and a \ninoone, falls within the $\PW\PZ$ topology.

Each SMS model is defined in terms of the initially produced pair of SUSY particles, here the \PSGc or $\slep$, and their decay chains, leading to the observable topologies.  For the interpretations we make assumptions for the production cross sections and branching fractions, motivated by predictions from theoretical models of soft SUSY breaking.  A review of these predictions for the electroweakino sector can be found in Ref.~\cite{Canepa:2020ntc}.  In these models the mixing matrix relating the gauge to mass eigenstates is approximately diagonal, so that we may identify the \PSGc states as bino-like, wino-like, and higgsino-like multiplets.  The mass splitting within these multiplets is small, while the hierarchy among the multiplets is dependent on the SUSY-breaking picture assumed.  For example, in ``natural'' SUSY models~\cite{Feng:1999zg,Hall:2011aa,Baer:2012up}, the higgsino-like multiplet tends to lie lowest among the \PSGc states.  As a representative of this class we consider the GMSB model~\cite{Dine:1981gu,Nappi:1982hm,Alvarez-Gaume:1981abe,Dine:1993yw,Dine:1994vc,Dine:1995ag,Giudice:1998bp,Matchev:1999ft}.
Spectra with a bino-like LSP are favored by alternative approaches to naturalness~\cite{Baer:2015tva} among other models.  The  next-to-lightest SUSY particle (NLSP) states can be either wino-like, with the higgsino-like states decoupled, or vice-versa; we consider both of these cases below.  

The following subsections provide detailed descriptions of the models:  three for electroweakino production and decay, and one that instead assumes that only the slepton states are accessible.  Of these, the higgsino-bino and slepton interpretations are in addition to those considered in the previous CMS combination paper~\cite{Sirunyan:2018ubx}.

\subsection{Models for the production of electroweakinos}

We consider three SMS models for the production and decay of electroweakinos in which the NLSP decays to the LSP with the emission of a \PW, \PZ, or \PH boson.  Here ``\PH boson'' refers to the observed $125\GeV$ scalar (Higgs) boson, assumed to be the lightest $CP$-even state of the extended Higgs sector.

\begin{figure}[!ht]
  \centering
  \includegraphics[width=0.33\textwidth]{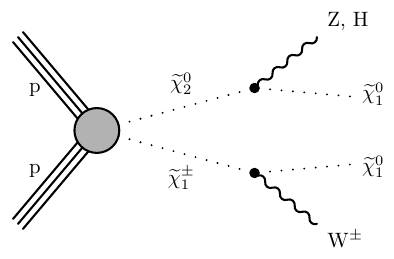}
  \caption{Wino-bino model: production of \ninotwo and \chinoone , with the \ninotwo decaying to either a \PZ or \PH boson and a \ninoone, and the \chinoone decaying to a \PW boson and a \ninoone.}
  \label{fig:sigWX}
\end{figure}

The first of these models is a wino-bino model,
Fig.~\ref{fig:sigWX}, in which the lightest chargino \chinoone is produced in association with the next-to-lightest neutralino \ninotwo.
In computing the production cross section we take the \chinoone and \ninotwo to be degenerate in mass and wino-like.
We consider mass splittings between the NLSP and LSP up to the kinematic limit of the experiment, but also include a search, ``$2/3\Pell$ soft''~\cite{CMS:2021edw}, targeting the compressed limit in which
the gaugino mass spectrum is approximately degenerate (gaugino mass universality~\cite{Canepa:2020ntc}).
In addition, allowing also for higgsino-gaugino mixing of the neutral states, and in the spirit of the more general SMS approach, we search for both $\PZ\ninoone$ and $\PH\ninoone$ decays of the $\ninotwo$, as indicated in Fig.~\ref{fig:sigWX}.

The other two electroweakino production models are characterized by a SUSY mass spectrum in which the light higgsinos, including the NLSP, are nearly degenerate in mass and heavier than the LSP.
In these models multiple production mechanisms leading to the same final state enhance the total effective cross section.

\begin{figure}[!ht]
  \centering
    \includegraphics[width=0.32\textwidth]{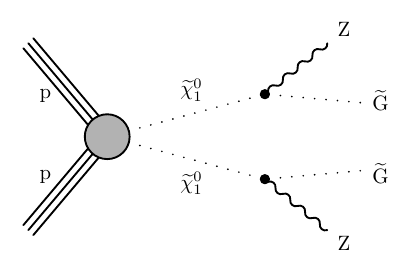}
    \includegraphics[width=0.32\textwidth]{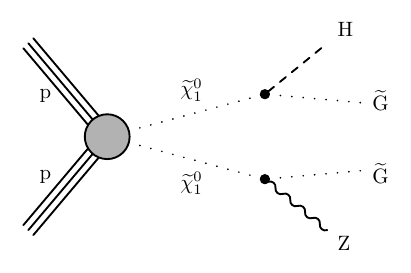}
    \includegraphics[width=0.32\textwidth]{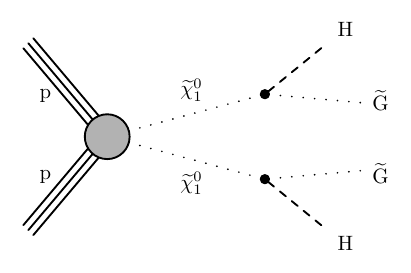}
  \caption{GMSB model: pair production of $\ninoone\ninoone$. The \ninoone particles each decay to a \gravitino with the emission of an SM gauge boson: (\cmsLeft) both \PZ, (middle) one \PZ and the other \PH, and (\cmsRight) both \PH. Soft fermions from decays of nearly degenerate neutralinos and charginos are omitted from these diagrams.}
  \label{fig:sigNeutral}
\end{figure}

The first of these, the GMSB model, is motivated by a specific GMSB scenario~\cite{Matchev:1999ft} in which the light higgsinos are nearly degenerate in mass, and the \gravitino is the LSP.  The coupling of \gravitino is suppressed by the SUSY breaking scale~\cite{Dimopoulos:1996vz}, so that the \ninoone NLSP is metastable.
We assume that nonetheless the \ninoone decays promptly on the measurement scale.
Thus the production of \ninoone pairs occurs both directly and as a result of cascade decays following production of the combinations $\chinoone\ninoone$, $\chinoone\ninotwo$, $\chinoone\chinoonemp$, and $\ninoone\ninotwo$.  Because of the small mass splitting among the higgsinos, low-momentum SM fermions emitted in these decays have a negligible impact on the kinematics of the event.  We consider in total three topologies as illustrated in Fig.~\ref{fig:sigNeutral}.  The \PZ decays (Fig.~~\ref{fig:sigNeutral}, left) of higgsino-like \PSGc states are included in the search topology, allowing for any admixture of gauginos into these states.

\begin{figure}[!ht]
  \centering
    \includegraphics[width=0.32\textwidth]{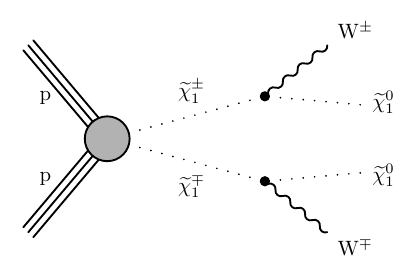}
    \includegraphics[width=0.32\textwidth]{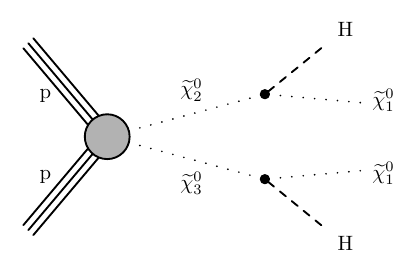}
    \includegraphics[width=0.32\textwidth]{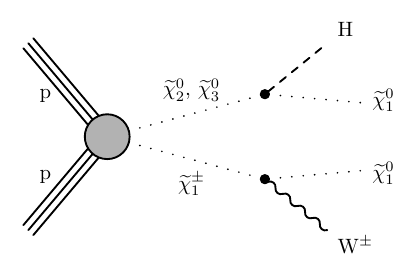}
  \caption{Higgsino-bino model: (\cmsLeft) the production of a pair of charginos followed by their decays to \PW bosons and the LSP, (middle) the production of a pair of neutralinos followed by decays to \PH bosons and the LSP, and (\cmsRight) the production of chargino-neutralino pairs followed by decay of the chargino (neutralino) to a \PW (\PH) boson and the LSP.}
  \label{fig:higgsino_bino}
\end{figure}

The second higgsino model, the higgsino-bino model, assumes a bino-like LSP, and again mass-degenerate light higgsinos. 
The production mechanisms considered are: $\chinoone\ninotwo$, $\chinoone\ninothree$, $\chinoone\chinoonemp$, and $\ninotwo\ninothree$.
The charged higgsinos decay to a \PW boson and the LSP with 100\% branching fraction, and for this process we assume that the neutral ones decay exclusively to {\PH}+LSP.
This scenario is motivated by an alternate approach to natural SUSY~\cite{Baer:2015tva} that seeks to evade constraints on the superpartner masses from earlier LHC searches by considering relatively large wino-bino mass splittings.
The topologies contributing the higgsino-bino interpretation are shown in Fig.~\ref{fig:higgsino_bino}. 

\subsection{Model for the production of sleptons}
\label{sec:sleptons}

\begin{figure}[!ht]
  \centering
    \includegraphics[width=0.33\textwidth]{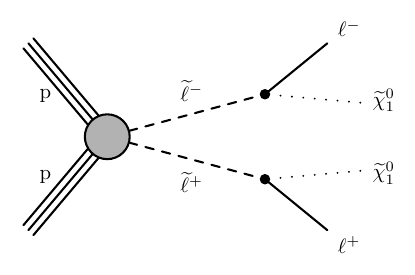}
  \caption{Slepton-neutralino model: direct slepton pair production, with each slepton decaying into a lepton and a \ninoone LSP.}
  \label{fig:sigSlepton}
\end{figure}
We also target the production of a pair of charged sleptons~\cite{ewkxsec1,ewkxsec5,ewkxsec7}, each of which decays to a lepton and a \ninoone LSP. 
The diagram for the process described by the slepton-neutralino model is shown in Fig.~\ref{fig:sigSlepton}, and the model assumes equal masses for the
superpartners $\PSe_{\text{L,\,R}}$ and $\PSGm_{\text{L,\,R}}$ of the electrons and muons, respectively, while the \PSGt states are considered to be decoupled.  
A dedicated analysis, presented here for the first time, addresses the compressed region of model parameter space that was not covered in a previous slepton search~\cite{Sirunyan:2020eab} by the CMS Collaboration.
 
\subsection{Search strategies}
\label{sec:search_strategies}
Here we give an overview of the strategies pursued by the analyses that are included in this combination.  Each is the subject of a more detailed summary given in Section~\ref{sec:searches}.  The contributing searches are:
\begin{itemize}
\item ``$2/3\Pell$ soft''~\cite{CMS:2021edw}, targeting the region of SUSY parameter space in which the mass difference between the NLSP and LSP is small.  At least one pair of opposite-sign, low-transverse momentum (\pt) electrons or muons (``light leptons'') is required, together with jets.
\item ``$2\Pell$''~\cite{Sirunyan:2020eab}, requiring two opposite-sign, same-flavor (OSSF) light leptons.  We consider two categories:
  \begin{itemize}
  \item ``$2\Pell$ on-\PZ'', in which the pair mass is consistent with the leptons' origin in the decay of a \PZ boson, and
  \item ``$2\Pell$ nonresonant'', requiring that the lepton pair mass lie outside the \PZ mass peak region.
  \end{itemize}
\item \multilep~\cite{CMS:2021cox}, selecting either two same-sign light leptons or at least three leptons, which may include one or more hadronically decaying tau leptons.
\item ``$1\Pell2\PQb$''~\cite{CMS:2021few}, requiring a light lepton consistent with arising from the decay of a \PW boson, and two \PQb quarks that form an \PH boson candidate.
\item ``$4\PQb$''~\cite{CMS:2022vpy}, requiring two resolved or merged pairs of \PQb-tagged jets that form \PH boson candidates.
\item ``Hadr.\ WX''~\cite{CMS:2022sfi}, selecting events with at least two merged (large-radius) jets from hadronically decaying \PW, \PZ, and \PH bosons.
\end{itemize}
In each of these searches a threshold on missing transverse momentum (\ptmiss) is imposed, since the presence of undetected LSPs in the final state is characteristic of the SUSY models considered.

Extensions are presented in this paper of the ``$2/3\Pell$ soft'' analysis reported in Ref.~\cite{CMS:2021edw}:  a more optimal binning of the search regions (SRs) is applied to the $\PW\PZ$ topology, and a new variable tailored to search for slepton pair production is exploited. Adjustments are made to the SR and control-region (CR) definitions of the \multilep analysis to remove overlap with the event selection of the ``$2/3\Pell$ soft'' analysis.

The combination is performed with a simultaneous maximum likelihood (ML) fit to the SR and CR data from the input searches, described in Section~\ref{sec:interpretations}.
In reference to background yields, we use the term ``post-fit'' to refer to this fit for the background-only hypothesis with all SRs and CRs included.

\section{The CMS detector}
\label{sec:detector}
The central feature of the CMS apparatus is a superconducting solenoid, 6\unit{m} in internal diameter and 13\unit{m} in length, that provides a magnetic field of 3.8\unit{T}. 
Within the solenoid volume are various particle detection systems. Charged-particle trajectories are measured by the silicon 
pixel and strip trackers.
A lead tungstate crystal electromagnetic calorimeter (ECAL) and a brass and scintillator hadron calorimeter (HCAL) surround the tracker volume, 
each composed of a barrel and two endcap sections. 
Forward calorimeters extend the pseudorapidity coverage provided by the barrel and endcap detectors. 
The calorimeters provide measurements of the energies of electrons, photons, and hadronic jets, as well as of the directions of jets.
Muons are measured in gas-ionization detectors embedded throughout the steel flux-return yoke outside the solenoid.  
A more detailed description of the CMS detector, together with a definition of the coordinate system used and the relevant kinematic variables, can be found in Ref.~\cite{Chatrchyan:2008zzk}.

Events of interest are selected using a two-tiered trigger system. The first level, composed of custom hardware processors, uses information from the calorimeters and muon detectors to select events at a rate of around 100\unit{kHz} within a fixed latency of about 4\mus~\cite{CMS:2020cmk}. The second, or high-level, trigger consists of a farm of processors running a version of the full event reconstruction software optimized for fast processing, and reduces the event rate to around 1\unit{kHz} before data storage~\cite{CMS:2016ngn}. 

\section{Event reconstruction and Monte Carlo simulation}
\label{sec:EventRecoMC}
All analyses considered as an input to this combination share common event reconstruction methods that are summarized below together with a general description of Monte Carlo (MC) simulated samples. Each analysis, however, targets a unique final state and different parts of kinematic phase space, and aims to maximize the background rejection.
As a result, the details of the selection, such as those concerning object identification and isolation, can vary from analysis to analysis to ensure the best sensitivity for a given search. These details,
in addition to the event selections, triggers, and MC simulated samples, are described in Refs.~\cite{Sirunyan:2020eab,CMS:2021few,CMS:2021edw,CMS:2021cox, CMS:2022vpy, CMS:2022sfi}.

\subsection{Common event reconstruction methods}
The particle-flow (PF) algorithm~\cite{CMS:2017yfk} aims to reconstruct and identify each individual particle in an event, with an optimized combination of information from the various elements of the CMS detector.  The energy of photons is obtained from the ECAL measurement. The energy of electrons is determined from a combination of the electron momentum at the primary interaction vertex as determined by the tracker, the energy of the corresponding ECAL cluster, and the energy sum of all bremsstrahlung photons spatially compatible with originating from the electron track. The energy of muons is obtained from the curvature of the corresponding track. The energy of charged hadrons is determined from a combination of their momentum measured in the tracker and the matching ECAL and HCAL energy deposits, corrected for the response function of the calorimeters to hadronic showers. Finally, the energy of neutral hadrons is obtained from the corresponding corrected ECAL and HCAL energies.

For each event, hadronic jets are clustered from PF candidates using the infrared- and collinear-safe anti-\kt algorithm~\cite{Cacciari:2008gp, Cacciari:2011ma}. The clustering is performed with distance parameter $R=0.4$ (AK4 jets) to reconstruct jets designed to contain the fragmentation products of a single parton, and with $R=0.8$ (AK8 jets) for jets originating from a multiparton system. Jet momentum is determined as the vectorial sum of the momenta of all reconstructed particles in the jet, and is found from simulation to be, on average, within 5--10\% of the true momentum over the whole \pt spectrum and detector acceptance. Besides the triggering event, \pp interactions within the same or nearby bunch crossings (pileup) contribute tracks and calorimetric energy depositions to the jet momentum. To mitigate this effect, charged particles identified to be originating from pileup vertices are discarded, and an offset term is applied to correct for remaining contributions to the energy of the jet.  Corrections derived from simulation are applied to match, on average, the measured response of jets to that of particle level jets. In situ measurements of the momentum balance in dijet, {photon}+{jet}, {\PZ}+jet, and multijet events are used to account for any residual differences in the jet energy scale between data and simulation~\cite{CMS:2016lmd}. The jet energy resolution amounts typically to 15--20\% at 30\GeV, 10\% at 100\GeV, and 5\% at 1\TeV~\cite{CMS:2016lmd}. Additional selection criteria are applied to remove jets potentially dominated by anomalous contributions from various subdetector components or reconstruction failures.

The primary vertex is taken to be the vertex corresponding to the hardest scattering in the event, evaluated using tracking information alone, as described in Section 9.4.1 of Ref.~\cite{CMS-TDR-15-02}.

The missing transverse momentum vector \ptvecmiss is computed as the negative vector sum of the transverse momenta of all the PF candidates in an event, and its magnitude is denoted by \ptmiss~\cite{CMS:2019ctu}. The \ptvecmiss is modified to account for corrections to the energy scale of the reconstructed jets in the event.

The identification of jets originating from {\PQb} quarks ({\PQb} tagging) is performed separately for AK4 and AK8 jets.
The AK4 \PQb jets are tagged with a version of the combined secondary vertex algorithm based on deep neural networks (\textsc{DeepCSV}~\cite{Sirunyan:2017ezt}), and 
the analyses entering this combination utilize various working points. 
The medium working point, which is common among multiple analyses, corresponds to an efficiency of about 68\% for a mistagging rate for light flavor quark and gluon jets of approximately 1\%.
The identification of AK8 jets containing two {\PQb} quarks is performed
with the deep-learning-based double-\PQb tagging algorithm \textsc{DeepDoubleBvL}~\cite{CMS-PAS-BTV-22-001,CMS-DP-2018-046}
with adversarial training~\cite{Louppe:2016ylz} to decorrelate the neural network tagging score and the jet mass.
AK8 jets are tagged as \PW, \PZ, or \PH bosons by the \textsc{DeepAK8} algorithm~\cite{CMS-PAS-BTV-22-001,CMS:2020poo}.

Hadronically decaying tau lepton candidates (\tauh) are reconstructed from PF candidates with
the ``hadron-plus-strips" algorithm~\cite{Khachatryan:2062435}. To reject the background originating from hadrons that are misreconstructed as \tauh, a boosted decision tree discriminant based on information from the reconstructed \tauh isolation, its measured lifetime, and the shape of the resulting jet is used~\cite{Sirunyan:2018pgf}. A strict selection criterion is applied in this discriminant, resulting on an efficiency of 50\% for selecting true \tauh decays and a less than 0.2\% mistagging rate.

\subsection{Monte Carlo simulation}
Signal samples are generated with {\MGvATNLO}~(2.2.2 or newer)~\cite{Alwall:2014hca,Frederix:2012ps} at leading-order (LO) precision, including up to two additional partons in the matrix element calculations. The production cross sections are
determined with approximate next-to-leading order (NLO) plus next-to-leading logarithmic (NLL)
accuracy~\cite{Beenakker:1999xh,ewkxsec1,ewkxsec2,Fuks:2012qx,Fuks:2013vua,ewkxsec5,ewkxsec6,ewkxsec7}, and are used to normalize the signal samples. For all signal samples, the lowest NLSP mass considered is 100\GeV, while the highest varies between 1000 and 1500\GeV depending on the model. While the LSP mass is fixed at 1\GeV in the GMSB higgsino model, it spans between 1 and 600\,(650)\GeV for the production of electroweakinos (sleptons). The mass points are generated with either 25 or 50\GeV steps in the uncompressed mass spectra scenarios, while for the compressed phase space a finer grid is used.

The background simulations rely on {\MGvATNLO}~(2.2.2 or newer)~, \POWHEG~(v1.0 or v2.0)~\cite{powheg0,powheg1,powheg2}, or \MCFM~7.0~\cite{mcfm0,mcfm1,mcfm2}, with calculations of either LO or NLO precision, all details depending on the process, as described in the individual published searches~\cite{Sirunyan:2020eab,CMS:2021few,CMS:2021edw,CMS:2021cox, CMS:2022vpy, CMS:2022sfi}.

Showering, hadronization, and the underlying event description are carried out using the \PYTHIA package 8.2~\cite{Sjostrand:2014zea} (the specific version number depends on the considered sample). The CUETP8M1 underlying event tune~\cite{Khachatryan:2015pea} is used for the SM background and signal generation with 2016 data-taking conditions. For the 2017 and 2018 samples, the CP5 and CP2 tunes~\cite{CMS:2019csb} are used for the SM background (NLO) and signal (LO) samples, respectively. Simulated samples generated at LO (NLO) with the CUETP8M1 tune use the parton distribution functions denoted by NNPDF3.0LO (NNPDF3.0NLO)~\cite{NNPDF:2014otw}, while those with the CP2 or CP5 tune use the NNPDF3.1LO (NNPDF3.1NNLO)~\cite{Ball:2017nwa} ones.
For all Monte Carlo samples, simulated minimum bias interactions are superimposed on the generated events, with a number distribution that is adjusted to match
the pileup distribution measured in data.
For some analyses, because scans over numerous mass points are required for the signal models, the response of the detector is described using the CMS fast simulation program~\cite{Abdullin:2011zz,Giammanco:2014bza},
which yields results that are generally
consistent with those from the simulation based on {\GEANTfour}~\cite{Agostinelli:2002hh} and have appropriate
systematic uncertainties applied to cover differences.

All simulated events are reconstructed with the same programs as those used for collision data.
Corrections are applied to simulated samples to account for differences between
the trigger, \PQb tagging, and lepton and photon efficiencies measured in data and the
\GEANTfour simulation. Additional differences arising from the use of fast simulation modeling,
such as selection efficiencies and the modeling of \ptmiss,
are corrected for and included in the systematic uncertainties.

\section{Individual analyses contributing to the combination}
\label{sec:searches}

The searches included in the combination are briefly described in this section, together with a more detailed account of any update made to an analysis since its publication. 

\subsection{Search in final states with two or three soft leptons and \texorpdfstring{\ptmiss}{ptmiss}}
\label{sec:SOS_description}

The ``$2/3\Pell$ soft'' search~\cite{CMS:2021edw} selects events with two or three low-\pt light leptons.  It targets the $\PW\PZ$ topology of the wino-bino model with chargino-neutralino production in the compressed region, i.e., where the mass difference \dm between \ninoone and the NLSP is small, such that the \PW and \PZ bosons are off-shell, and the observable decay products have low momentum. Events are divided into two search categories, ``$2\Pell$ soft'' and ``$3\Pell$ soft''.  Both require an $\Pep\Pem$ or $\PGmp\PGmm$ pair, jets, and large \ptmiss.  The presence of an additional electron or muon defines the ``$3\Pell$ soft''category.
The leptons are required to satisfy $\pt < 30\GeV$, with the minimum value of \pt being 3.5 (5.0)\GeV for muons (electrons), except as noted in the following paragraph. 

The variables used to discriminate signal from background are the dilepton mass \mll, \ptmiss, and \ptmisscorr; the latter is defined as the magnitude of the total \ptvecmiss reconstructed excluding the muon transverse momentum vectors, as an approximation to the trigger-level \ptmiss quantity used during data-taking. 
Events in the two search categories are binned in \ptmisscorr as detailed in Table~\ref{tab:SOS_MET_bins}, resulting in four (two) SRs in the ``$2\Pell$ soft'' (``$3\Pell$ soft'') category. 
Additionally, in the low-$\ptmiss$ SRs all events must satisfy $\ptmiss>125\GeV$, and only muon pairs are accepted because of the utilized triggers; the minimum value of \pt is increased to 5\GeV for these muons to ensure efficiency of the trigger.  
The trigger paths targeting the low-$\ptmiss$ SRs were disabled during some parts of the data taking, resulting in the slightly lower integrated luminosity of 129\fbinv. 
Each \ptmisscorr bin is further divided into regions of \mll with boundaries of 1, 4, 10, 20, 30, and 50\GeV, with the first bin removed in the low-$\ptmiss$ category because of the trigger requirements.
For the ``$2\Pell$ soft'' category \mll is the dilepton invariant mass $\text{M}(\Pell\Pell)$, whereas for the ``$3\Pell$ soft'' category it is the minimum of the invariant masses of the possible OSSF pairs, $\text{M}^{\text{min}}_{\text{OSSF}}(\Pell\Pell)$.
In the following, \mll is used to denote the dilepton mass in either category.

Contributions from $\PJGy$ and $\PGU$ mesons are removed by excluding events with \mll in the ranges 3.0--3.2 and 9.0--10.5\GeV, respectively. Additional kinematic requirements are applied to further reduce backgrounds.  The largest backgrounds arise from $\PZ/\gamma^*$ and $\cPqt\cPaqt$ production, as well as events with nonprompt leptons from mainly $\PW+\text{jets}$ ($2\Pell$) or \ttbar ($3\Pell$) production.

\begin{table}[!hbtp]
    \topcaption{``$2/3\Pell$ soft'' search: definition of the lepton multiplicity and \ptmisscorr SRs.  The boundaries are indicated in\GeVns. Events in the low-$\ptmiss$ SR must additionally have $\ptmiss>125\GeV$. 
    }
    \label{tab:SOS_MET_bins}
    \centering
    \begin{scotch}{c c c c c }
         SR & Low-$\ptmiss$ & Med-$\ptmiss$ & High-$\ptmiss$ & Ultrahigh-$\ptmiss$\\[\cmsTabSkip]
        & $\ptmisscorr$ & $\ptmisscorr$ & $\ptmisscorr$ & $\ptmisscorr$ \\
         $2\Pell$ soft & $[125,200]$ & $[200,240]$ & $[240,290]$ & $>$290 \\
         $3\Pell$ soft & $[125,200]$ & \multicolumn{3}{c}{$>$200} \\
    \end{scotch}
\end{table} 

Two major updates are made to the search~\cite{CMS:2021edw}, designed with no prior knowledge of the distribution of data events in the updated SRs, and discussed in the following subsections. The first one is an improvement to the signal extraction, while the second introduces a new search for slepton pair production.

\subsubsection{Improved signal extraction: parametric \texorpdfstring{$\dm$}{dm} binning}
\label{sec:ParametricSE}

\begin{figure}[!ht]
  \centering
  \includegraphics[width=0.47\textwidth]{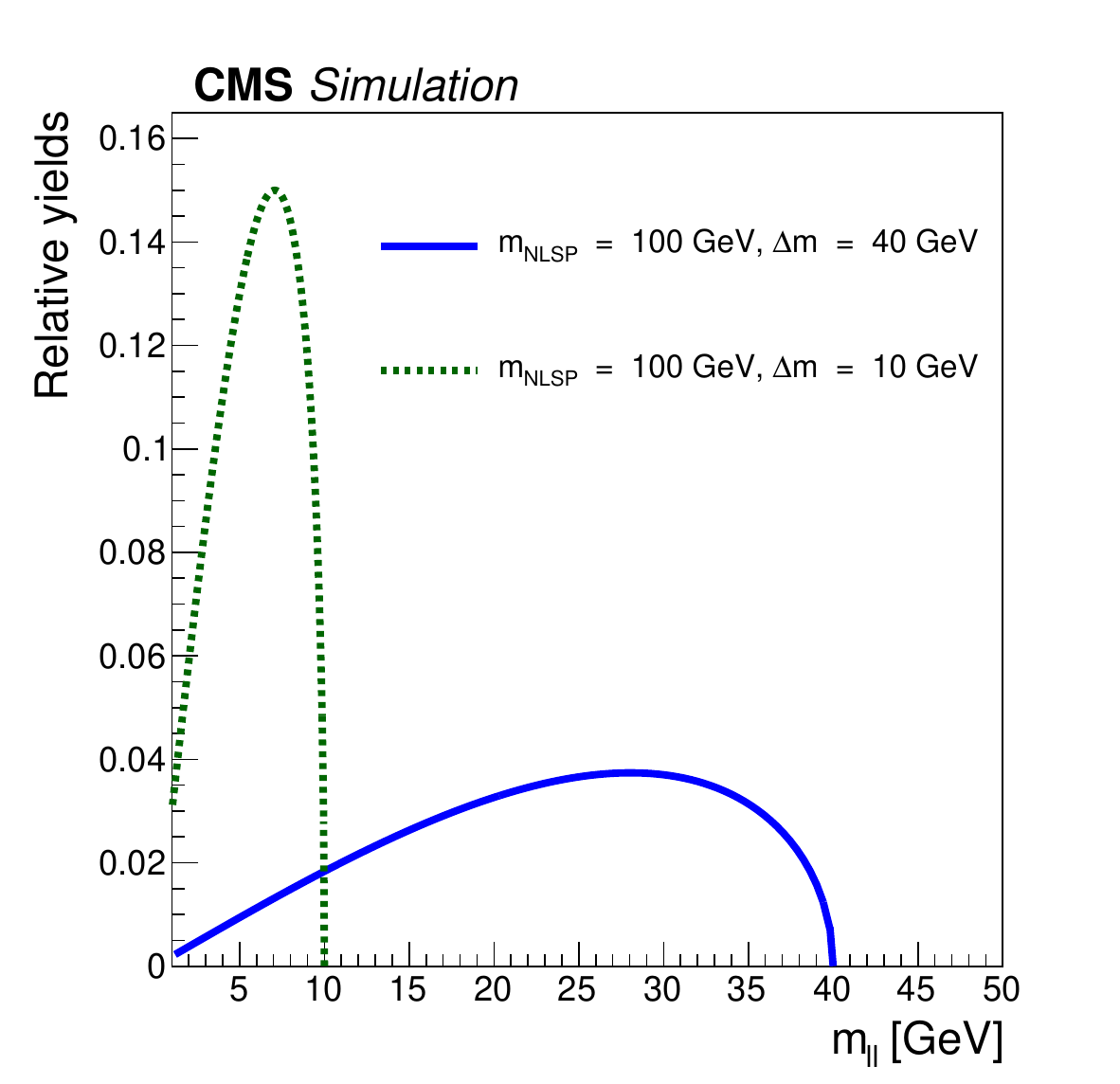}
  \caption{``$2/3\Pell$ soft'' search:  dilepton mass spectrum for two mass hypotheses with the same NLSP mass (100\GeV) and different mass splittings $\dm$ (40 or $10\GeV$), both corresponding to analytical phase space only calculations. The distributions have a kinematic endpoint at the mass splitting.}
  \label{fig:SOSTheoryShape1}
\end{figure}

For signal, \mll corresponds to the mass of the off-shell \PZ-boson $\PZ^*$ from the decay $\ninotwo \to \ninoone \PZ^*$. The \mll spectrum has a kinematic end-point at \dm, and the shape depends on the sign of the product of the two smallest (in magnitude) eigenvalues of the neutralino mass matrix~\cite{DeSanctis:2007yoa}. The dilepton mass spectra are corrected accordingly in Ref.~\cite{CMS:2021edw}, but for this combination we assume a flat phase space for the \ninotwo decay rate to ensure consistency among the input analyses. Two examples of the spectrum are shown in Fig.~\ref{fig:SOSTheoryShape1}.

To better exploit the shape of the \mll distribution, we optimize the \mll binning for each \dm hypothesis, separately for each lepton category and \ptmiss bin.
The procedure is iterative.
The range $[1\GeV, \dm]$ is first divided into four bins of equal signal quantiles.
A fifth bin spans $[\dm, 50\GeV]$, which is expected to be devoid of signal. Aside from resolution effects and statistical fluctuations, this construction leads to distributions of the events per bin that, for the bins below \dm, are uniform for signal, and depleted for background.

In the second step we check that none of the resulting bins introduces large statistical fluctuations of the expected SM background, 
taking into account both the relative importance of the various background processes and the statistical precision of their predictions.
This is to avoid an artificial increase in the sensitivity.
If the statistical fluctuations for any of the bins in $[1\GeV, \dm]$ exceed those reported in Ref.~\cite{CMS:2021edw}, 
the number of bins in that range is reduced by one, and the binning is rederived.
This process is repeated until a binning is found that satisfies this requirement,
or results in the minimal two-bin case $[1\GeV, \dm]$, $[\dm, 50\GeV]$.
The CR binning and background estimation strategy remain the same as in Ref.~\cite{CMS:2021edw}, ensuring that there are no changes to how the CRs are correlated with the SRs.

\begin{figure*}[!th]
  \centering
    \includegraphics[width=0.45\textwidth]{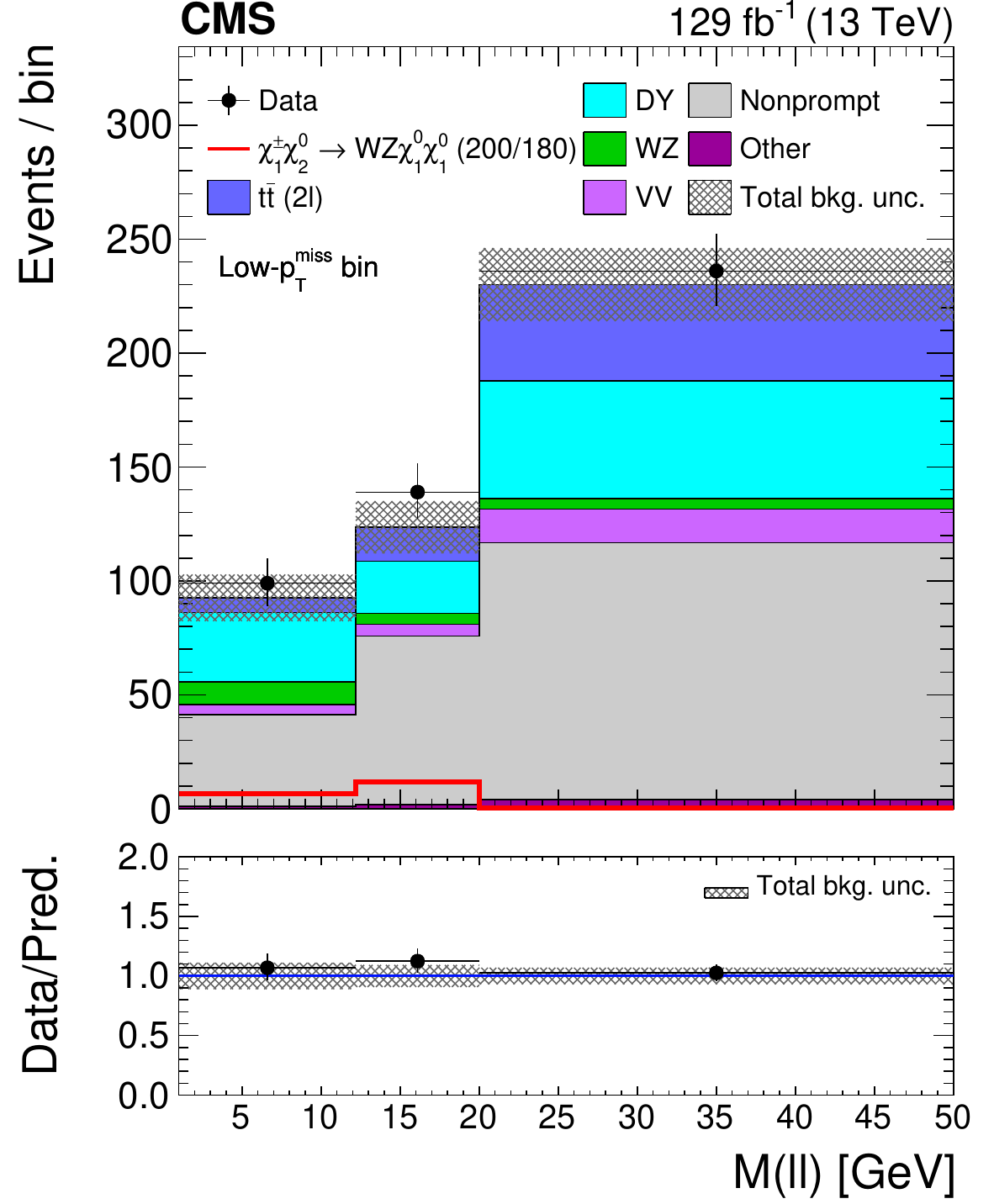}
    \includegraphics[width=0.45\textwidth]{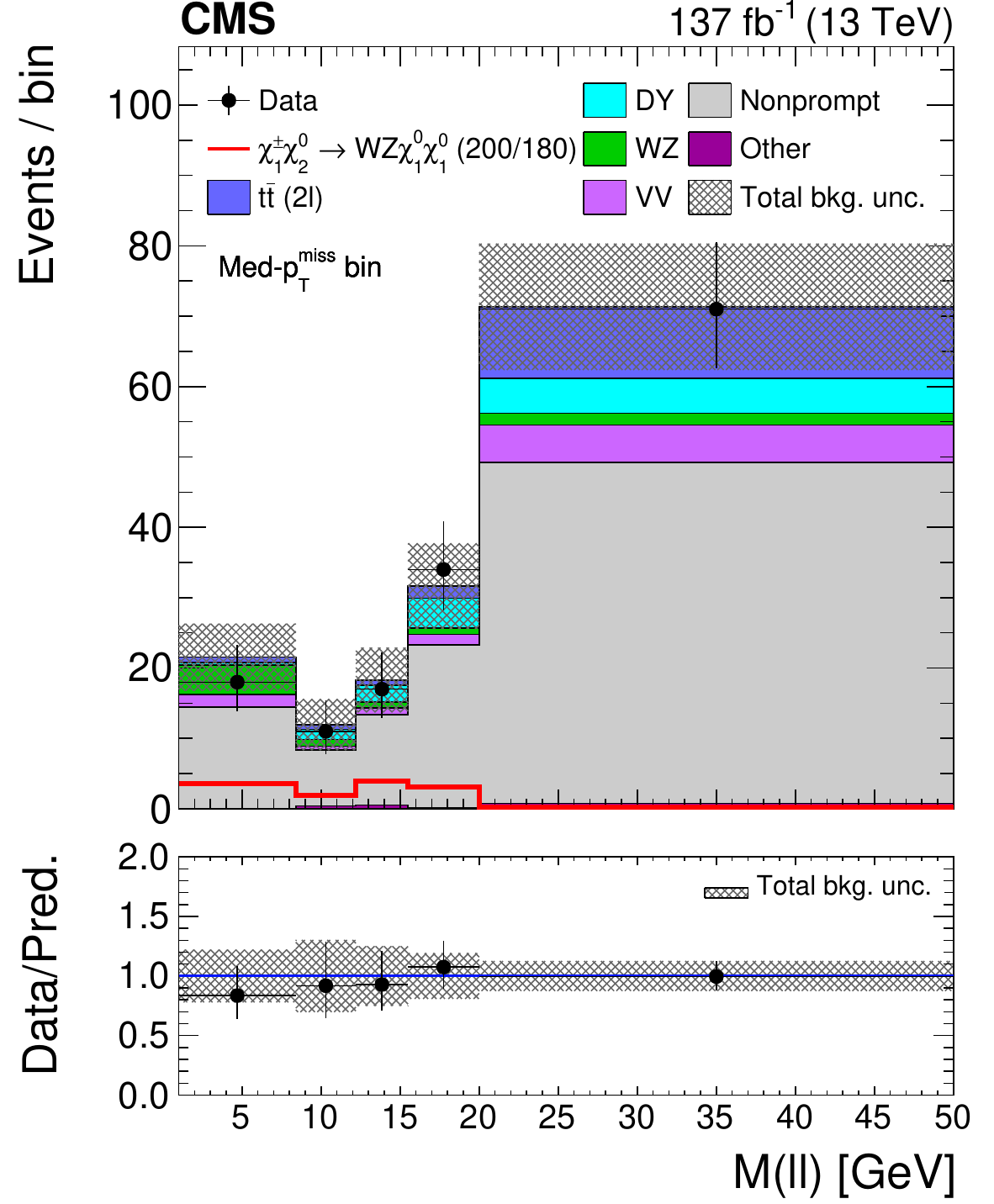}
    \includegraphics[width=0.45\textwidth]{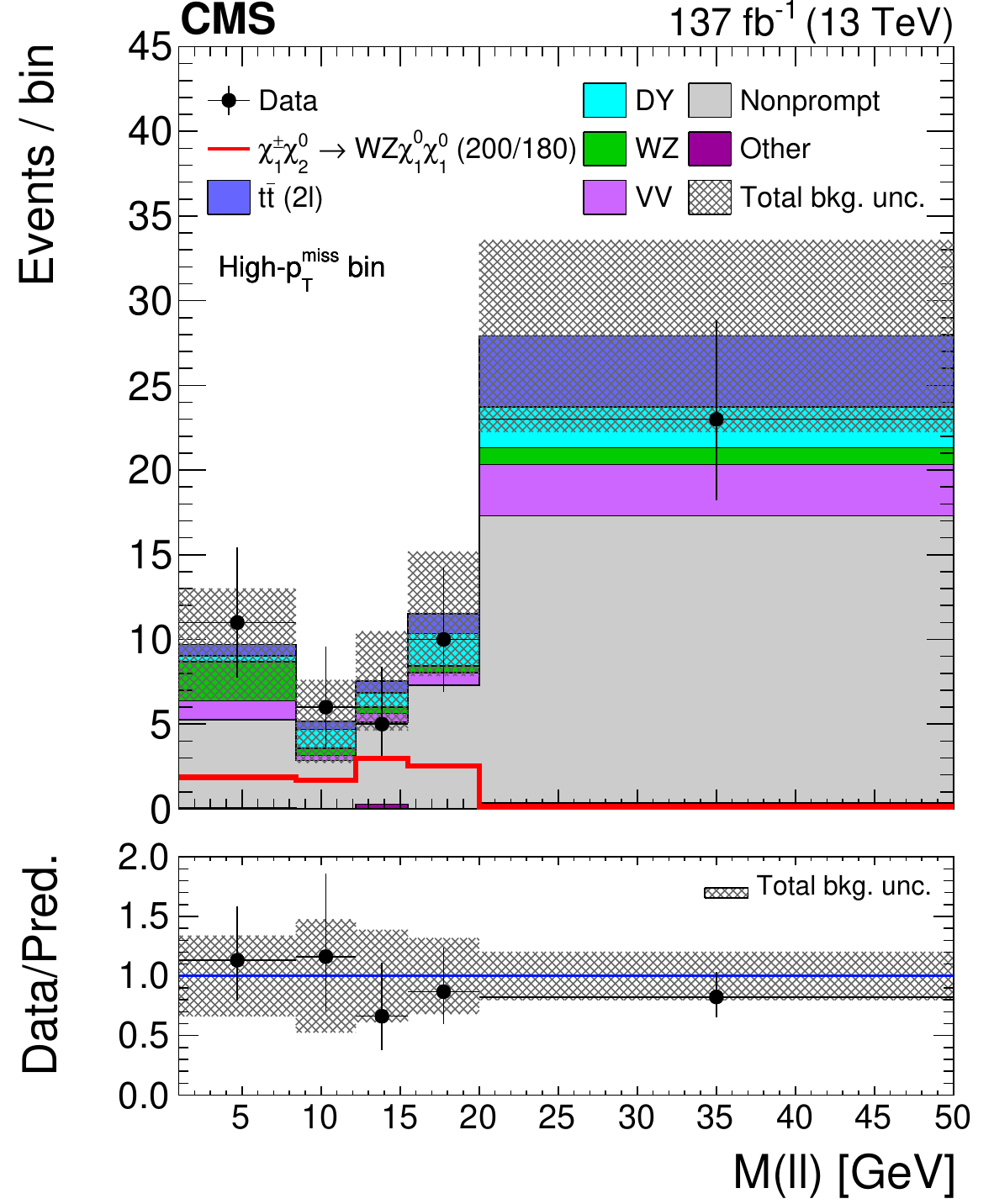}
    \includegraphics[width=0.45\textwidth]{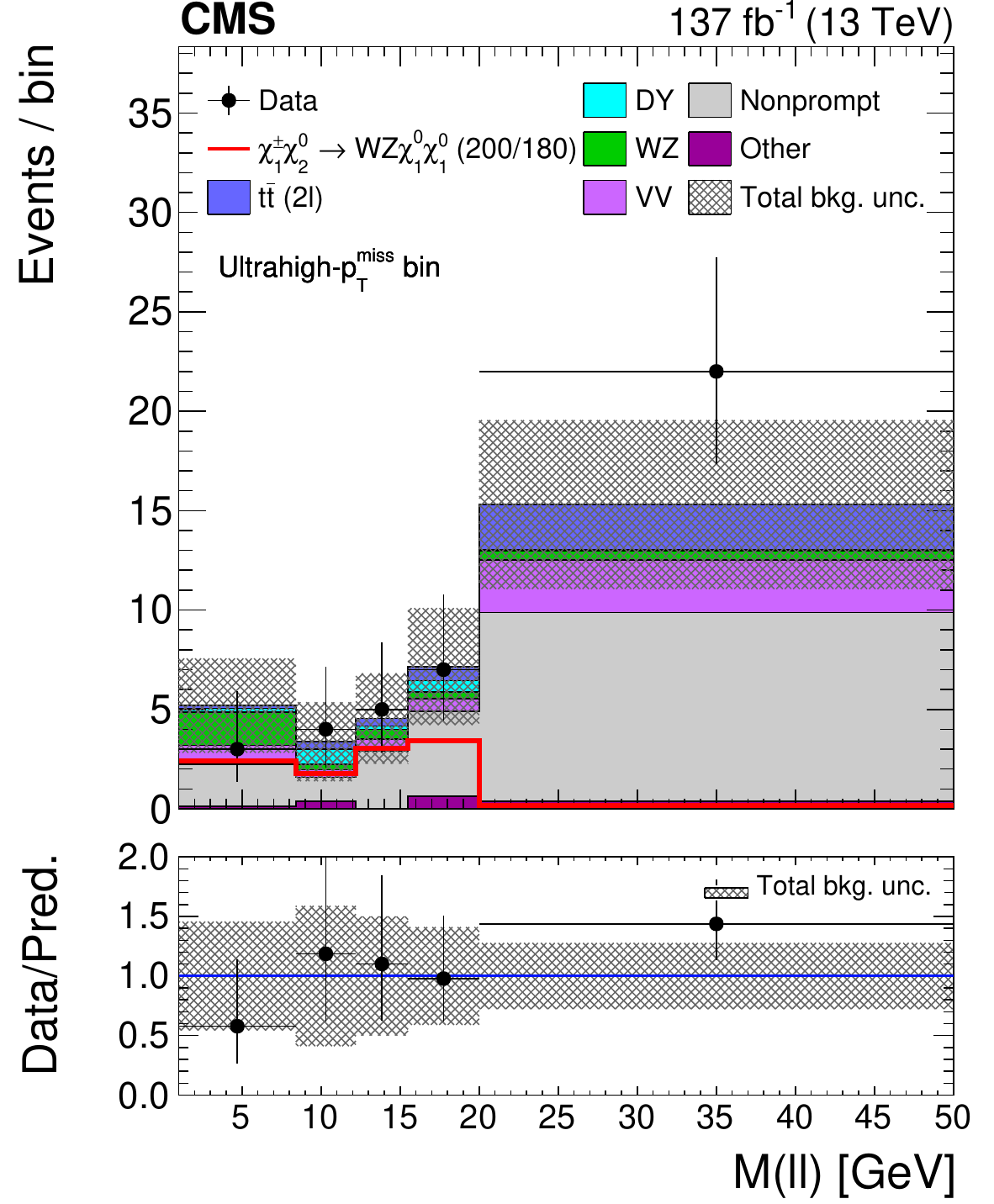}    
  \caption{``$2/3\Pell$ soft'' search: post-fit distributions of the $\text{M}(\Pell\Pell)$ variable for the low- (upper left), medium- (upper right), high- (lower left), and ultrahigh- (lower right) \ptmiss bins in the ``$2\Pell$ soft'' signal region of Ref.~\cite{CMS:2021edw}. These distributions are based on the parametric binnings derived for signal mass points with $\dm = 20\GeV$. The pre-fit signal distribution for $m_{\chinoone} = m_{\ninotwo}=200\GeV, m_{\nino}=180\GeV$ is overlaid for illustration. ``Nonprompt'' refers to the background contribution arising from nonprompt or misidentified leptons.}
  \label{fig:sos_tchiwz_2lSR}
\end{figure*}

\begin{figure*}[!ht]
  \centering
    \includegraphics[width=0.45\textwidth]{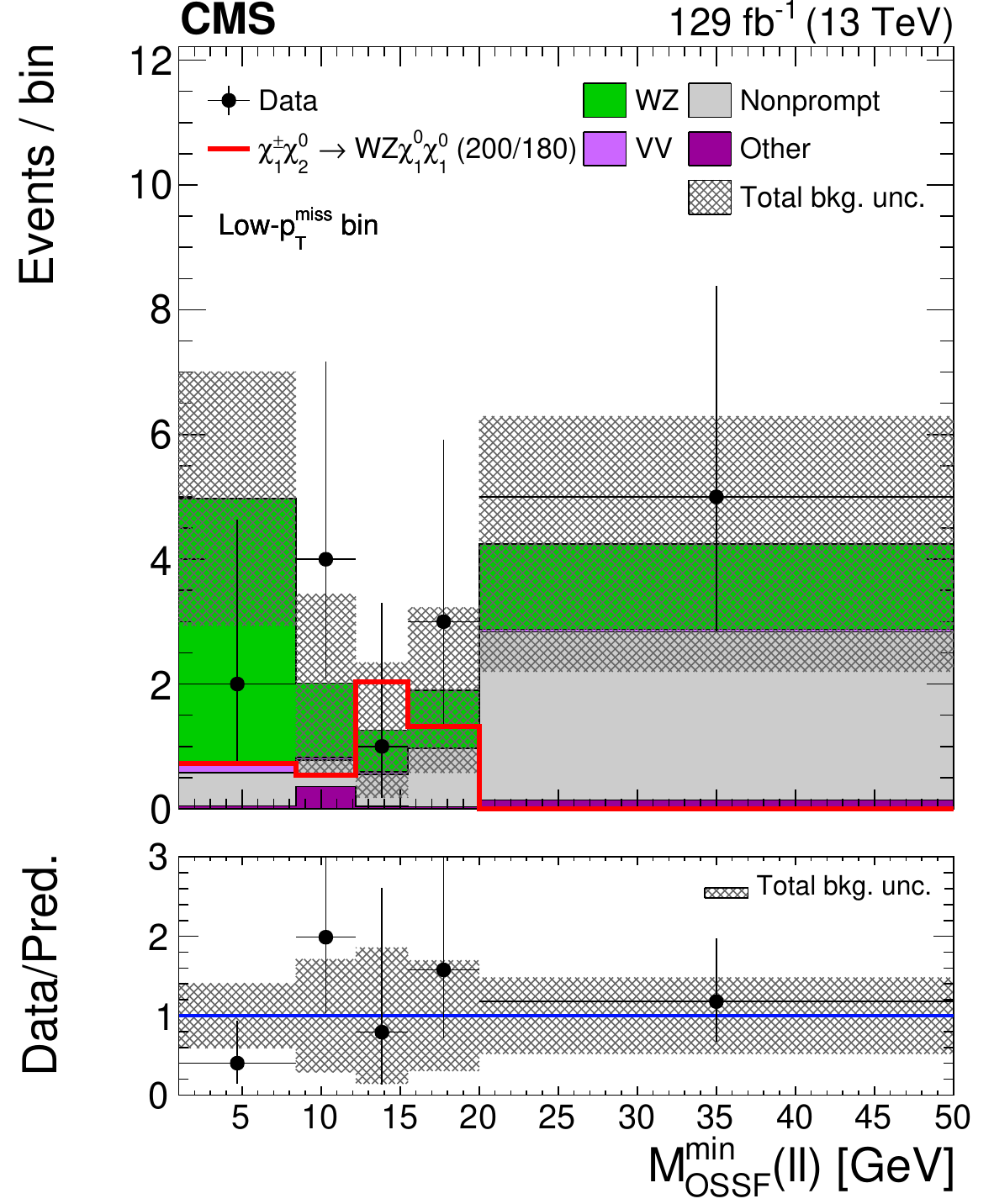}
    \includegraphics[width=0.45\textwidth]{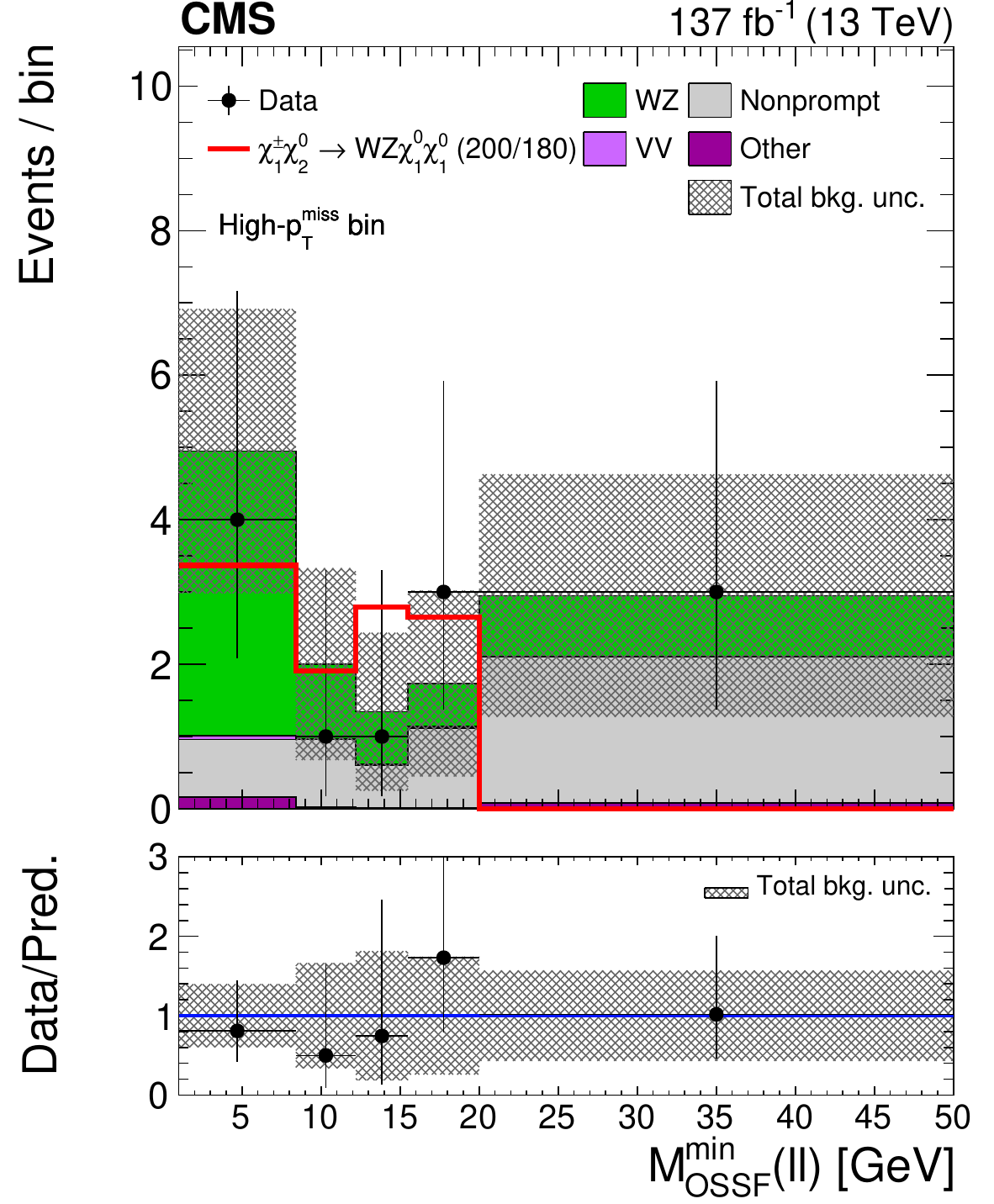}
  \caption{``$2/3\Pell$ soft'' search: post-fit distributions of the $\text{M}^{\text{min}}_{\text{OSSF}}(\Pell\Pell)$ variable for the low- (left) and medium- (right) \ptmiss bins in the ``$3\Pell$ soft'' signal region of Ref.~\cite{CMS:2021edw}. These distributions are based on the parametric binnings derived for signal mass points with $\dm = 20\GeV$. The pre-fit signal distribution for $m_{\chinoone} = m_{\ninotwo}=200\GeV, m_{\nino}=180\GeV$ is overlaid for illustration. ``Nonprompt'' refers to the background contribution arising from nonprompt or misidentified leptons.}
  \label{fig:sos_tchiwz_3lSR}
\end{figure*}

For each of the nine \dm hypotheses considered in this search, six parametric binnings have been derived (one for each SR). 
Examples of the resulting post-fit SR distributions of \mll with parametric binnings (for signal mass points with $\dm = 20\GeV$) are shown in Figs.~\ref{fig:sos_tchiwz_2lSR} and \ref{fig:sos_tchiwz_3lSR} for the ``$2\Pell$ soft'' and ``$3\Pell$ soft'' search categories, respectively.

The expected sensitivity of the search is improved over the full parameter space considered in this analysis. The largest gains are obtained for mass splittings of $\dm<20\GeV$, where the expected sensitivity to the mass of \ninotwo is increased by 5--25\GeV, depending on \dm.

\subsubsection{Slepton production in the compressed region}\label{sec:sosSleptons}

The production of a smuon or selectron pair for the case of a compressed SUSY spectrum has not been studied previously in a dedicated search by the CMS Collaboration.  
Here we analyze the events captured by the ``$2\Pell$ soft'' selection to search for this process, employing the parametric binning procedure described in Section~\ref{sec:ParametricSE} to optimize the sensitivity. 
In this case, to discriminate between signal and background we replace \mll with the kinematic variable $\mTii(\text{k}, \text{k}, \chi_{m_\chi})$~\cite{Lester:1999tx,Barr:2003rg}.
For a process producing two particles of equal mass $M$ that each decay into an invisible daughter $\chi$ of mass $m_\chi$ and a light visible object $\text{k}$, we have
\begin{linenomath}
\ifthenelse{\boolean{cms@external}}
{\begin{multline}
\mTii (\text{k}, \text{k}, \chi_{m_\chi}) = \min_{\ptvecmiss{}^{(1)} + \ptvecmiss{}^{(2)} = \ptvecmiss}
  \Big( \max \Big[\\
  \mt^{(1)}(m_\chi), \mt^{(2)}(m_\chi) \Big] \Big).
\label{eq:MT2}
\end{multline}
}{\begin{equation}
\mTii (\text{k}, \text{k}, \chi_{m_\chi}) = \min_{\ptvecmiss{}^{(1)} + \ptvecmiss{}^{(2)} = \ptvecmiss}
  \left(max \left[ \mt^{(1)}(m_\chi), \mt^{(2)}(m_\chi) \right] \right).
\label{eq:MT2}
\end{equation}
}
\end{linenomath}
The constraint enforces that the observed \ptvecmiss is equal to the sum of  $\ptvecmiss{}^{(i)}$ ($i=1,2$), two vectors in the transverse plane that represent the \pt vectors of the invisible objects. 
The $\mt^{(i)}$ are the transverse masses obtained by pairing the $\ptvecmiss{}^{(i)}$ with either of the two visible objects.  The distribution in $\mTii(\text{k}, \text{k}, \chi_{m_\chi})$ extends between $m_\chi$ and $M$. 
Here the visible objects are the two low-\pt leptons, and we take $m_\chi$ to be 100\GeV, noting that the exact choice of $m_\chi$ is expected to have a negligible effect on the sensitivity of the search. In the following, we utilize the notation $\mTii(\Pell, \Pell, \chi_{100})$.

\begin{figure*}[!ht]
\centering
        \includegraphics[width=0.45\textwidth]{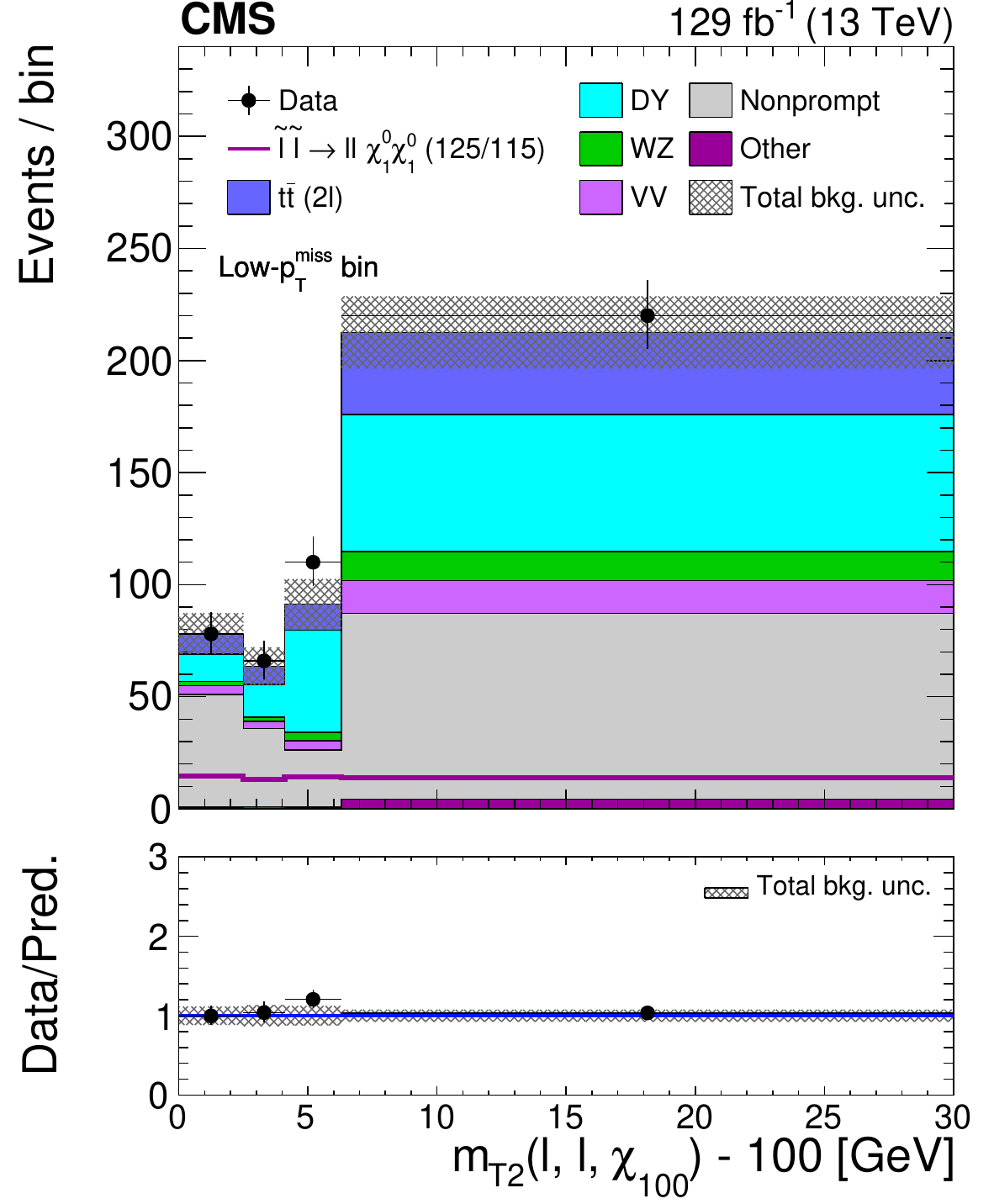}
        \includegraphics[width=0.45\textwidth]{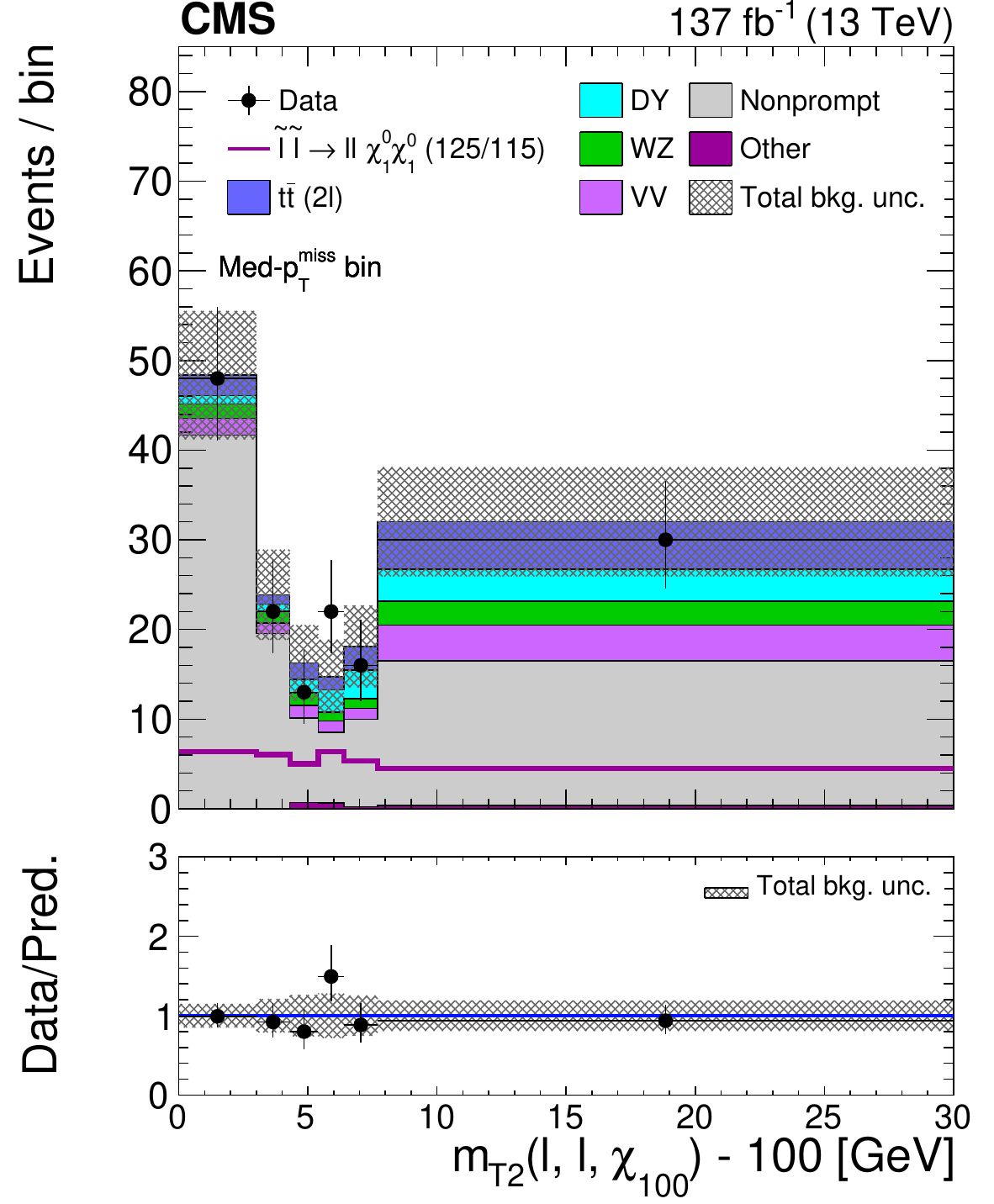}
        \includegraphics[width=0.45\textwidth]{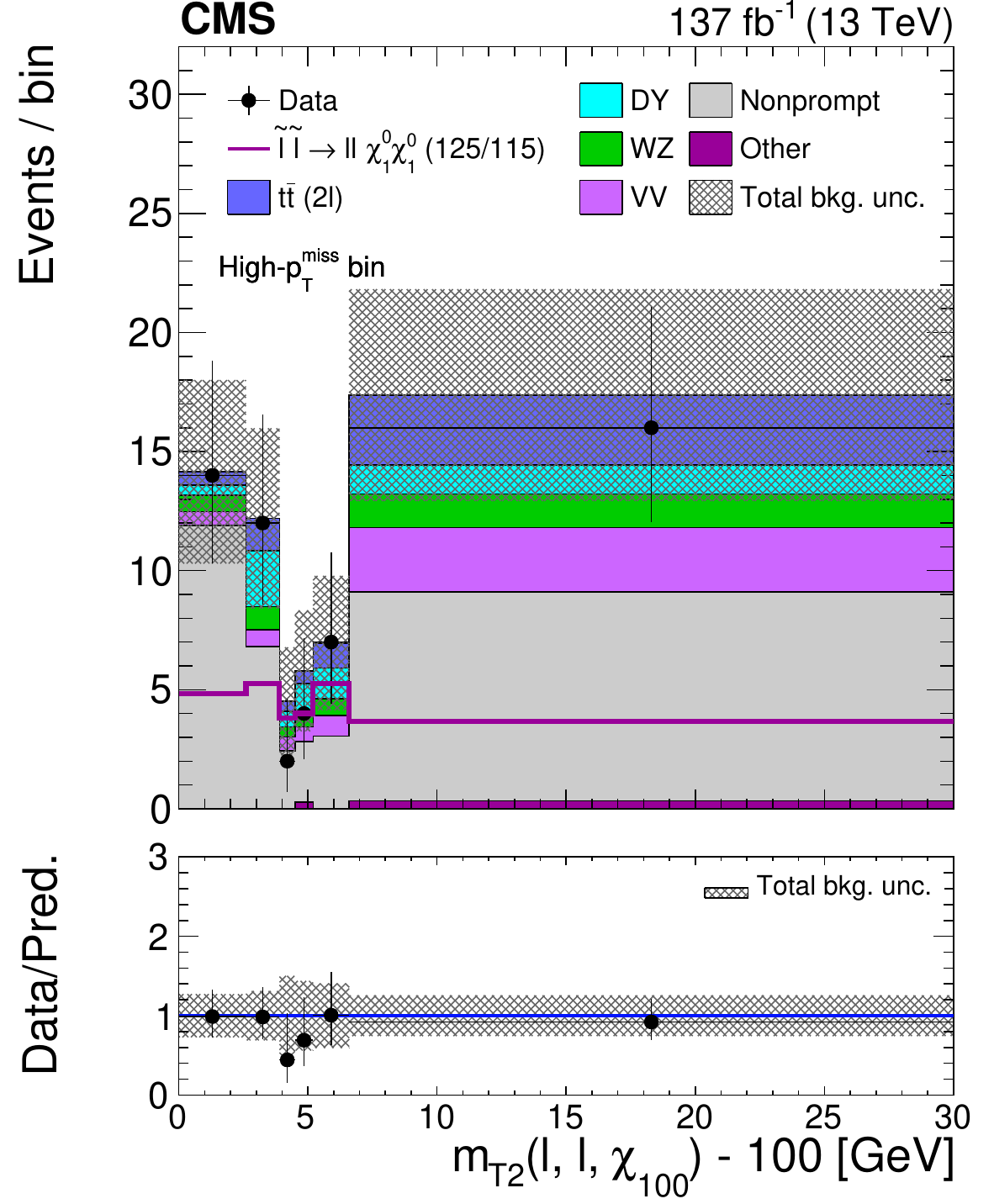}
        \includegraphics[width=0.45\textwidth]{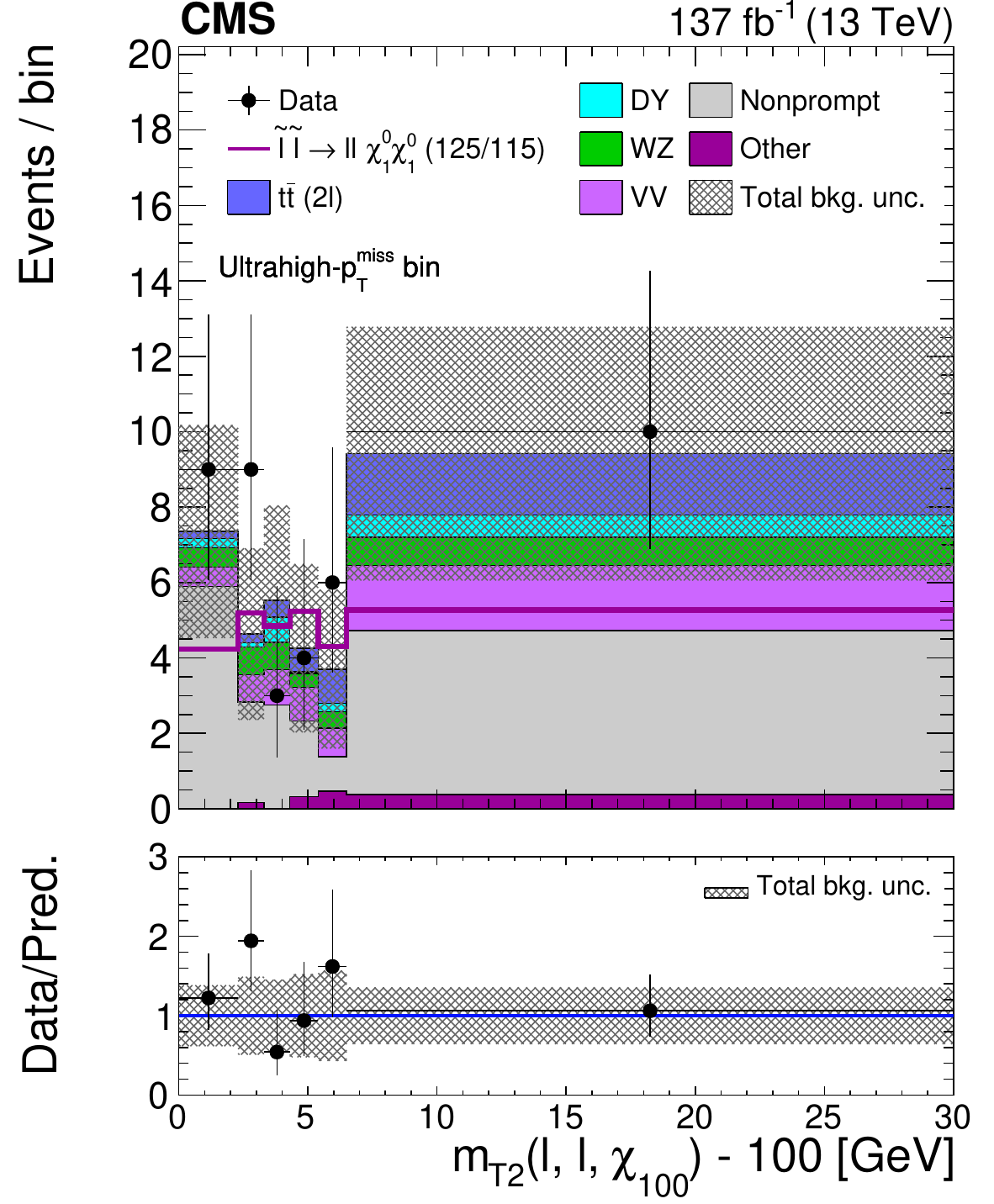}
\caption{``$2/3\Pell$ soft'' search: post-fit distributions of the $\mTii(\Pell, \Pell, \chi_{100})$ variable are shown for the low- (upper left), medium- (upper right), high- (lower left), and ultrahigh- (lower right) \ptmiss bins in the ``$2\Pell$ soft'' signal region of Ref.~\cite{CMS:2021edw}. These distributions are based on the parametric binnings derived for the mass-point $m_{\slep}=125\GeV, m_{\nino}=115\GeV$, for which the pre-fit signal distribution is overlaid for illustration. Note that the signal distribution (purple line) is approximately flat across $\mTii(\Pell, \Pell, \chi_{100})$, by construction of the parametric binning procedure. The minimum value of $\mTii(\Pell, \Pell, \chi_{100})$, $m_\chi = 100\GeV$, is subtracted for the abscissa of the plot. ``Nonprompt'' refers to the background contribution arising from nonprompt or misidentified leptons.}
\label{fig:sos_slepton_SR}
\end{figure*}

The analysis utilizes the same background estimation methods, CRs, and systematic uncertainties as Ref.~\cite{CMS:2021edw}, with the exclusion of the ``$3\Pell$ soft'' SRs. 
A separate parametric binning is chosen for each model point, $(m_{\slep},\dm)$. An example of the resulting post-fit SR distributions of $\mTii(\Pell, \Pell, \chi_{100})$, for mass point ($m_{\slep}=125\GeV, m_{\nino}=115\GeV)$, is shown in Fig.~\ref{fig:sos_slepton_SR}. The SM background normalizations are constrained in the same way as done for the search for charginos and neutralinos~\cite{CMS:2021edw}, including the \mll distribution in the CR.

\subsection{Searches in final states with an opposite-sign same-flavor lepton pair and \texorpdfstring{\ptmiss}{ptmiss}}
\label{sec:zedge_description}

These searches~\cite{Sirunyan:2020eab} select events with lepton pairs $\Pep\Pem$ or $\PGmp\PGmm$ and large \ptmiss.
Of the SRs defined in Ref.~\cite{Sirunyan:2020eab}, we include in the present combination those designated ``EW on-\PZ'' and ``Slepton''.  The first targets the wino-bino and GMSB models, while the second targets the slepton-neutralino model, both in the uncompressed region of the model parameter space.
Events in the two categories are required to have exactly two reconstructed OSSF leptons, with $\pt > 25$\,(20)\GeV for the leading (trailing)-\pt lepton.  Threshold requirements are also imposed on the event \ptmiss and $\mTii(\Pell, \Pell, \chi_{0})$, or $\mTii(\Pell\PQb, \Pell\PQb, \chi_{0})$ for SRs with two \PQb-tagged jets, to suppress SM backgrounds.

The main backgrounds in this final state are Drell--Yan lepton pairs with \ptmiss from mismeasured jet energies, \ttbar production, and events with both a \PZ boson and prompt neutrinos, including $\PQt\PAQt\PZ$, $\PW\PZ\to 3\Pell \PGn$, and $\PZ\PZ\to 2\Pell2\PGn$.  They are estimated with CRs in the data.

\subsubsection{\texorpdfstring{On-\PZ search (``$2\Pell$ on-\PZ'')}{On-Z search (``2l on-Z'')} }

The ``$2\Pell$ on-\PZ'' search focuses on cases in which the invariant mass \mll of the dilepton pair is consistent with the \PZ boson mass:  $86  < \mll < 96\GeV$. This targets the $\PW\PZ$, $\PZ\PZ$, and $\PH\PZ$ topologies, where the $\PZ$ boson (one of the $\PZ$ bosons in the $\PZ\PZ$ topology) is produced on-shell and decays to two electrons or muons, while the other boson decays hadronically.

In addition to the two leptons, we require the presence of jets produced in the hadronic decay of the bosons. A total of three SRs in this category are considered, of which two target the $\PW\PZ$ and $\PZ\PZ$ topologies, and one is dedicated to the $\PH\PZ$ topology. Events in the resolved $\PW\PZ$ and $\PZ\PZ$ regions are required to have at least two jets and no \PQb-tagged jets.  Events in the Lorentz-boosted region have $<$2 AK4 jets and at least one AK8 jet satisfying mass and structure requirements.  The $\PH\PZ$ SR requires two \PQb-tagged jets consistent with an $\PH\to \bbbar$ decay. The three SRs are then binned in \ptmiss. 

\subsubsection{\texorpdfstring{Off-\PZ search (``$2\Pell$ nonresonant'')}{Off-Z search ("2l nonresonant")}}

The ``$2\Pell$ nonresonant'' SRs target nonresonant production of sleptons. In this case,
the dilepton invariant mass must satisfy $20  < \mll < 65\GeV$, or $\mll > 120\GeV$. Two SRs (``Slepton'') are defined, those with and without jets in addition to the lepton pair.  These SRs are further separated into bins in \ptmiss.

\subsection{Search in final states with two same-sign leptons, or three or more leptons, and \texorpdfstring{\ptmiss}{ptmiss}}
\label{sec:ML_description}

The \multilep search~\cite{CMS:2021cox} focuses on final states with either two light same-sign leptons, or three or more leptons including \tauh, and \ptmiss.
The analysis targets the $\PW\PZ$, $\PW\PH$, $\PZ\PZ$, $\PH\PZ$, and $\PH\PH$ topologies.

The search is organized in a series of twelve exclusive categories that target the different decay modes of the \PW, \PZ, and \PH bosons.
The categories are defined based on lepton multiplicity, flavor, and charge~\cite{CMS:2021cox}, summarized for purposes of this paper as follows:
\begin{itemize}
    \item Category SS: a pair of same-sign light leptons and no \tauh.
    \item Category A: three light leptons, including an OSSF lepton pair.
    \item Category B: three light leptons, with no OSSF lepton pairs.
    \item Categories C, D, E, and F: three leptons, at least one of which is a \tauh
    \item Categories G, H, I, J, and K: more than three leptons.
\end{itemize}

Within each category, several SRs are defined by requirements on multiple kinematical variables, including particle \pt,
$\mTii(\Pell, \Pell, \chi_{0})$, \mll, and 
the scalar \pt sum \HT for all selected jets with $\pt>30\GeV$, as detailed in Ref.~\cite{CMS:2021cox}. 
For category A, an alternative approach based on parametric neural networks is used to build discriminant variables for the $\PW\PZ$ model~\cite{CMS:2021cox}.

For the SS category, the main sources of background 
are nonprompt leptons or charge mismeasurements, estimated from control samples in data.
Those for the other light-lepton categories A, B, G, and H result from $\PW\PZ$ and $\PZ\PZ$ production, and are estimated using MC simulation that is validated in low-\ptmiss CRs.
For the remaining categories, those that contain \tauh candidates,
the main backgrounds arise from Drell--Yan and \ttbar production with associated nonprompt \tauh candidates.
The predictions for these contributions are obtained directly from data using a loose-to-tight selection method~\cite{CMS:2021cox}. 

\paragraph*{Modifications to remove overlap with the ``$2/3\Pell$ soft'' analysis}~\\[-15pt]

In Ref.~\cite{CMS:2021cox}, the requirements on the minimum \pt of electrons and muons vary between 10 and 25\GeV, chosen to ensure full trigger efficiency after the selection.  Because of a small overlap between this and the ``$2/3\Pell$ soft'' analysis, for the present combination these threshold \pt values were raised to $30\GeV$ for the leading lepton in categories A and B.  In this section we describe the changes in SR yields resulting from this modification.

The first consequence of the altered selection is a slight revision in the definitions of the CRs used for evaluation of the backgrounds from $\PW\PZ$ production and from asymmetric photon-to-electron conversions. The first of these CRs is included in all the analysis fits, while the second is used to validate the MC simulation of the conversion background.
The level of agreement between the observed and expected yields in the CRs that was found in the previously published analysis is maintained.
The effect of the tightened lepton \pt requirement on the $\PW\PZ$ CR is nearly negligible; around 1\% of the overall background events are lost. As photon-to-electron conversions tend to result in electrons of lower momenta, the effects of restricting the leading-lepton \pt are greater in the conversion CR: around 10\% of the overall CR yield is lost. 

\begin{figure*}[!ht]
\centering
  \includegraphics[width=\textwidth]{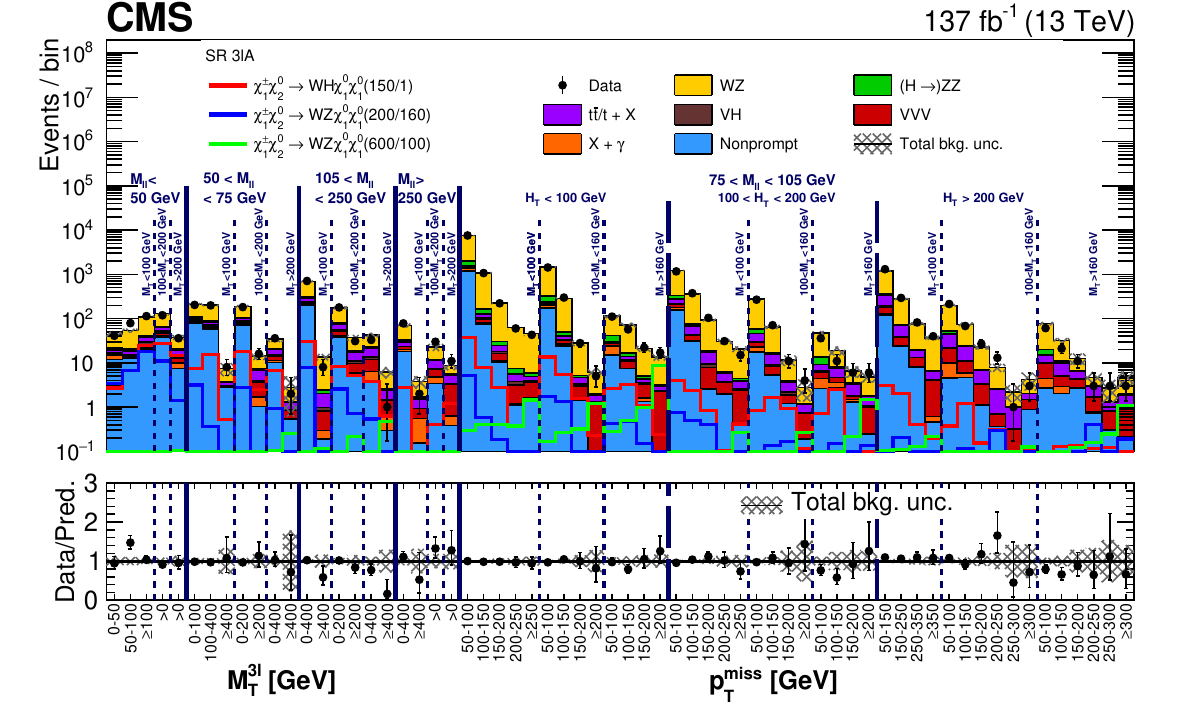}
\caption{\multilep search: observed and expected event yields across the SRs in category A, events with three light leptons of which at least two form an OSSF pair, after the requirement that the leading-lepton \pt be greater than 30\GeV is applied. The pre-fit signal distributions for three mass-points are overlaid for illustration, and the considered mass hypothesis is indicated in the legend with notation
$(m(\text{NLSP})/m(\text{LSP}))$; for example (150/1) stands for $m_{\chinoone} = m_{\ninotwo}=150\GeV, m_{\nino}=1\GeV$. ``Nonprompt'' refers to the background contribution arising from nonprompt or misidentified leptons.}
\label{fig:ML_3lApt30}
\end{figure*}

The change in the minimum leading-lepton \pt also introduces slight variations in the signal extraction strategy, as not all SRs are affected equally by the tighter criteria. The SR definitions themselves are unchanged: those in category A are divided according to the value of \mll of the OSSF lepton pair into on-\PZ, with $75 < \mll < 105\GeV$, and off-\PZ subcategories.
Each of these subcategories is further divided into bins in \ptmiss, \HT,
the transverse mass (\mt) of \ptmiss and the lepton not in the OSSF pair, and the transverse mass (\mtlll) of the three leptons and \ptmiss. The exact definitions of the SRs of category A are shown
by the labeling in Fig.~\ref{fig:ML_3lApt30}, which gives the distribution of yields in those SRs with the altered leading-lepton \pt threshold.

The additional requirement causes a reduction in the yields of the overall SM backgrounds; for most SRs the effect on these yields is less than 1\%, except for regions with low invariant mass ($\mll < 75\GeV$) where the decrease in the background yields can be up to 7\%. For the signal models the decrease in acceptance is strongly correlated with the model parameters. Signal models that target uncompressed scenarios are largely unaffected, while those that include compressed spectra have a reduction in overall yields of 5--7\%.

The evaluation of the neural network shapes for the $\PW\PZ$ model of category A is kept unchanged after the additional leading-lepton \pt selection is applied. 

\begin{figure}[!ht]
\centering
  \includegraphics[width=0.45\textwidth]{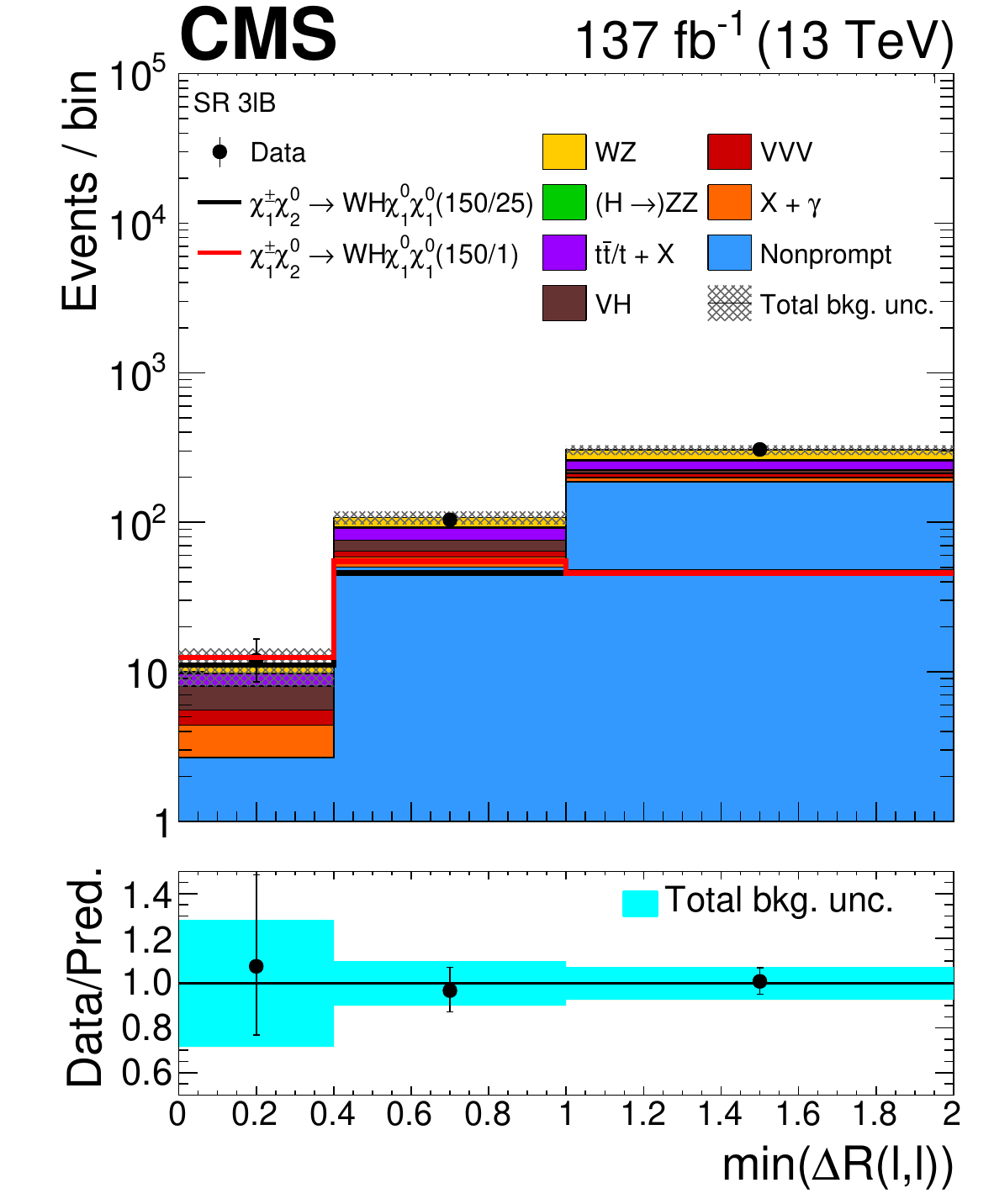}
\caption{\multilep search: observed and expected event yields across the SRs in category B, events with three light leptons and no OSSF pair, after the requirement that the leading-lepton \pt be greater than 30\GeV is applied. The pre-fit signal distributions for two mass-points are overlaid for illustration, and the considered mass hypothesis is indicated in the legend with notation
$(m(\text{NLSP})/m(\text{LSP}))$; for example (150/1) stands for $m_{\chinoone} = m_{\ninotwo}=150\GeV, m_{\nino}=1\GeV$. ``Nonprompt'' refers to the background contribution arising from nonprompt or misidentified leptons.}
\label{fig:ML_3lBpt30}
\end{figure}

In category B, the SR definition strategy is much simpler, as both the available event yields and number of studied models are smaller than in category A. In this case, the signal extraction strategy is based on a single variable, the minimum distance $\min \Delta R(\Pell,\Pell)$ between any two leptons. Figure \ref{fig:ML_3lBpt30} shows the expected and observed distributions of events in this category after the additional requirement is applied. The SM background contributions are reduced by roughly 5\%, with the largest changes in the signal-depleted second and third bins. The signal efficiency is reduced by 1--2\% overall.

The effect of the additional lepton \pt requirement on the analysis performance has been studied in terms of the expected upper limits for the GMSB model with degenerate higgsinos (leading to the topologies $\PH\PH$, $\PH\PZ$, and $\PZ\PZ$), and for the wino-bino model with production of \chinoone and \ninotwo ($\PW\PZ$, $\PW\PH$, and mixed-topology models). In most cases the upper limits are consistent at the 1\% level before and after the additional requirement is applied. The only exceptions are the compressed scenarios in the $\PW\PZ$ model, where decreases of 1--10\% in expected sensitivity to the mass of NLSP are seen between mass splittings of $\dm=70$ and $20\GeV$. The effect translates to a loss of roughly 10\GeV in NLSP mass exclusion in the worst case scenario.
In the combined analysis, this loss of sensitivity is compensated by the performance of the ``$2/3\Pell$ soft'' lepton analysis.

\subsection{Search in final states with one lepton, two \texorpdfstring{\PQb}{b} jets, and \texorpdfstring{\ptmiss}{ptmiss}}
The ``$1\Pell2\PQb$'' search~\cite{CMS:2021few} targets the $\PW\PH$ topology by focusing on events with one electron or muon ($\pt > 30\GeV$) produced in a \PW boson decay, two \PQb-tagged jets from a Higgs boson decay, and large \ptmiss from the two LSPs and the neutrino.  The search benefits from the large branching fraction for $\PH \to \PQb\PAQb$ ($58\%$)~\cite{ParticleDataGroup:2022pth} and the suppression of multijet background in the leptonic final state.

The two \PQb-tagged jets  are required to have an invariant mass consistent with that of the Higgs boson (90--150\GeV). 
The transverse mass of the lepton and \ptmiss is required to be larger than 150\GeV to remove events where the lepton and the \ptmiss both arise from a single leptonic \PW decay. 
In addition, events where the Higgs bosons are Lorentz boosted are selected by requiring a single large-radius jet ($R=0.8$), identified by the \textsc{DeepAK8} algorithm~\cite{CMS:2020poo}. To enhance the sensitivity,
the selected events are sorted into SRs according to the presence or not of a boosted \PH-tagged jet, \ptmiss, and the number of AK4 jets.

The main backgrounds for the search arise from top quark production and \PW boson production. The contributions of these processes to the SR yields are estimated using observed yields in CRs scaled by transfer factors obtained from simulated samples. These extrapolations are studied extensively in validation regions orthogonal to the SRs in order to assess the systematic uncertainty in the background measurement.

\subsection{Search in final states with four \texorpdfstring{\PQb}{b} jets and \texorpdfstring{\ptmiss}{ptmiss}}

The ``$4\PQb$'' search~\cite{CMS:2022vpy} selects events with \PQb-tagged hadronic jets, no charged leptons, and large \ptmiss, to target the HH topology through the decays $\PH\to\bbbar$. 
In each event, we consider either a resolved signature, in which the two \PQb quarks of each \PH decay are contained in two separate AK4 jets, or a boosted signature, where the two \PQb quarks are both contained in a single AK8 jet (``$\PQb\PQb$-tagged jet'').

For the SRs of the resolved signature at least four (and to suppress combinatorial background no more than five) AK4 jets with $\pt>30\GeV$ are required.  Separate SRs are defined for which three (3-\PQb) or at least four (4-\PQb) jets satisfy \PQb-tagging criteria.  
The two \PH candidates are chosen by considering the four jets with the highest \PQb tagging discriminator scores, forming all possible pairs of these jets, and selecting the pairing that minimizes the difference in the invariant masses 
of the two pairs.  This difference is then required to be less than 40\GeV, and the average $\langle m_{\PQb\PQb}\rangle$ of the invariant masses of the two pairs is required to lie within the \PH mass peak region, $100<\langle m_{\PQb\PQb}\rangle<140\GeV$. An upper limit is imposed on the largest angular separation \drmax between pairs of the \PH daughter jets to remove lepton+jets \ttbar events, the primary background to this search. 

For the boosted signature, at least two AK8 jets with $\pt>300\GeV$ are required, with SRs having one (``1-{\PQb}\PQb'') or at least two (``2-{\PQb}\PQb'') of these satisfying $\PQb\PQb$ tagging requirements. The \PH candidates are the two highest-\pt AK8 jets meeting these criteria.  Each of these jets is required to have a mass, as computed by the ``soft drop'' algorithm~\cite{Larkoski:2014wba,Dasgupta:2013ihk}, that lies in the \PH mass peak region, 60--260\GeV.
The 1-bb CRs and SRs of the boosted signature are excluded from the higgsino-bino interpretation to avoid overlap with the ``Hadr. WX'' search.

Both signatures' SRs are binned in \ptmiss, while for the resolved signature an additional binning in \drmax further increases the sensitivity. 

In addition to lepton+jets \ttbar  production, background contributions arise at higher \ptmiss from \PW or \PZ production in association with jets. The backgrounds from all sources are predicted using data CRs that require either exactly two \PQb jets (resolved) or no $\PQb\PAQb$ jets (boosted), or an \PH candidate mass outside the defined \PH mass peak region.

\subsection{Search in final states with multiple jets and \texorpdfstring{\ptmiss}{ptmiss}}
The ``Hadr.\ WX'' search~\cite{CMS:2022sfi} selects events with a pair of AK8 jets and large $\ptmiss$. The search targets the hadronically decaying $\PW\PW$, $\PW\PZ$, and $\PW\PH$ signal topologies.  We require at least 2 AK8 jets, $\ptmiss > 200\GeV$, $\HT > 300\GeV$, and no charged leptons.  To reduce backgrounds with higher jet multiplicities, we further require at most 6 AK4 jets.  (For signal, the requirement of two AK8 jets generally implies the presence of around four overlapping AK4 jets.)  Taking advantage of the larger hadronic branching fraction of \PH and \PZ bosons, in particular to pairs of bottom quarks, the search is classified broadly into two SRs.  The \PQb-veto SR requires zero \PQb-tagged jets, to be sensitive to the topologies $\PW\PW$, and $\PW\PZ$ with the \PZ boson decaying to hadronic final states without \PQb quarks.  The \PQb-tag SR requires at least one \PQb-tagged jet, for the topologies $\PW\PZ$ and $\PW\PH$ with \PZ or \PH decaying to \bbbar.  The \PQb-tag SR is further divided into one with a $\PQb\PQb$-tagged \PH candidate jet ({\PW}\PH) and one without (\PW).  A third \PQb-tag SR (\PH)~\cite{CMS:2022sfi}
without a specific requirement on \PW candidate jets
is excluded, together with the corresponding CR, from the higgsino-bino interpretation to avoid overlap with the ``4\PQb'' analysis.

The \textsc{DeepAK8} algorithm~\cite{CMS:2020poo} is used to tag the decays of the \PW, \PZ, and \PH bosons.
This algorithm provides multiple tagging scores that distinguish AK8 jets arising from different hadronic decay modes of the SM bosons and the top quark.
The algorithm also provides two versions of neural networks with and without adversarial training~\cite{Louppe:2016ylz} to decorrelate the neural network tagging score and the jet mass.
The mass-decorrelated version of the \textsc{DeepAK8} \PW tagger (referred to as the V tagger) is used to tag hadronic decays of both \PW and \PZ bosons,
while the \textsc{DeepAK8} \PW tagger without mass decorrelation (referred to as the \PW tagger) is used to tag decays of \PW bosons.
The V tagger provides good efficiency for \PZ as well as \PW jets, while the dedicated \PW tagger has a lower misidentification rate.
In the \PQb-veto search region we require that at least one AK8 jet be tagged by the \PW tagger, and at least one other AK8 jet be tagged by the V tagger.
The AK8 jet mass is required to be between 65 and 105\GeV.
In the \PQb-tag search region the \PW tagger and mass requirement are utilized for the AK8 jets from \PW bosons,
while the AK8 jets from \PZ and \PH bosons are tagged using the \textsc{DeepAK8} $\PQb\PQb$ tagger with mass decorrelation and with the jet mass requirement of 75--140\GeV.

The main backgrounds in this search include $\PW$+jets, $\PZ$+jets, and $\PQt\PAQt$ processes.
These backgrounds are estimated with techniques based on CRs in data that are disjoint from the corresponding SRs.
The CRs are defined by inverting the tagging requirements for the AK8 jets or by requiring one charged electron or muon candidate.
Transfer factors from the CRs to the SRs are derived from simulation and multiplied by the CR yields to obtain the estimated background yields in the SRs.

\section{Combination strategy}
\label{sec:combStrat}

A simultaneous maximum likelihood fit to the signal and control regions of the searches described above is made for each signal model, as detailed in Section~\ref{sec:interpretations}.
Table~\ref{table:all_searches} shows which SRs are used for each model and topology.
It should be noted that not all analyses target the same signal parameter space. For example, the ``$2/3\Pell$ soft'' lepton analysis has sensitivity for compressed signals due to the low-\pt leptons, while the rest target the mostly uncompressed signals.

\begin{table*}[!ht]
  \topcaption{Summary of the searches considered in the combination and the
    SRs that contribute to the interpretation of each signal model and topology.
The following abbreviations appear in the table: 
For the ``$2\Pell$ on-\PZ'' analysis, ``EW'' refers to the resolved and boosted VZ SRs and the HZ SR.  The row label ``$2\Pell$ nonres.''\ refers to the ``$2\Pell$ nonresonant'' search, and in that row ``Slepton'' refers to the two dedicated slepton SRs, those requiring $N_{\text{jet}}=0$ and $N_{\text{jet}}>0$.  For the \multilep search, ``A(NN)'' indicates SR A with the parametric neural network signal extraction.  For the "Hadr. WX'', ``ex H'' means all SRs except the \PQb-tag \PH SR.}
\label{table:all_searches}
\cmsTable{
\begin{scotch}{ l  | c c | c c c | c c c | c}
  & \multicolumn{2}{c|}{Wino-bino} & \multicolumn{3}{c|}{GMSB} & \multicolumn{3}{c|}{Higgsino-bino} & Sleptons \\
Search & \wz & \wh & \zz & \zh & \hh & \ww & \hh & \wh & $\Pell^+\Pell^-$ \\
$2/3\Pell$ soft~\cite{CMS:2021edw} & all &  &  &  &  &  &  &  & $2\Pell$ soft \\
$2\Pell$ on-\PZ~\cite{Sirunyan:2020eab} & EW &  & EW & EW &  &  &  &  & \\
$2\Pell$ nonres.~\cite{Sirunyan:2020eab}  &  &  &  &  &  &  &  &  & Slepton \\[4pt]
\multirow{2}{*}{\multilepNoQuotes~\cite{CMS:2021cox}} & SS, & SS, & \multirow{2}{*}{all} & \multirow{2}{*}{all} & \multirow{2}{*}{all} &  & & SS, &  \\
  & A(NN) & A--F &  &  &  &  &  & A--F &  \\[4pt]
$1\Pell2\PQb$~\cite{CMS:2021few} &   & all &  &  & &  &  & all  &  \\[4pt]
\multirow{2}{*}{$4\PQb$~\cite{CMS:2022vpy}} &  &  &  &  &
\multirow{2}{*}{all} &  & 3-\PQb, 4-\PQb, &  &  \\
  &  &  &  &  &  &  & 2-{\PQb}\PQb &  &  \\[4pt]
Hadr. WX~\cite{CMS:2022sfi} & all & all &  &  &  & ex H &  & ex H &  \\
\end{scotch}
}
\end{table*}

The SRs and CRs are those defined in the original publications, except as noted in Section~\ref{sec:searches}.  The modifications made to avoid sample overlap are briefly summarized here.
For the \multilep analysis a small change was made to the leading-lepton $\pt$ selection in categories A and B to avoid overlaps with the ``$2/3\Pell$\ 
soft'' search.  The ``3$\Pell$'' \wz CR of the ``$2/3\Pell$ soft'' search was removed to avoid overlap with the aforementioned categories of the \multilep analysis when these searches are combined.  To provide a constraint on the background contribution for the ``$2/3\Pell$ soft'' lepton search in the combined fit, the $\PW\PZ$ backgrounds of the two analyses are correlated by utilizing a common nuisance parameter in the fit, constrained by a region of the \multilep analysis. 
For the higgsino-bino combination, to avoid overlap between the ``Hadr.~\wx'' and ``$4\PQb$'' searches, the \PQb-tag \PH SRs and CRs of the ``Hadr.~\wx'' analysis and the single $\PQb\PQb$-tag SRs and CRs of the boosted ``$4\PQb$'' sample are removed.

\section{Systematic uncertainties}
\label{sec:syst}

Systematic uncertainties in each of the input analyses are given in the respective publications~\cite{Sirunyan:2020eab,CMS:2021few,CMS:2021edw,CMS:2021cox, CMS:2022vpy, CMS:2022sfi}.
The sources and their level of correlation across the analyses are shown in Table~\ref{table:systsCondensed}. Some analyses have additional uncertainties beyond these, which are analysis-specific and treated as uncorrelated.

The signal yield predictions are affected by the experimental calibration uncertainties listed in Table~\ref{table:systsCondensed}.  Uncertainties are also computed for theoretical and experimental inputs to the simulations, specifically the renormalization and factorization scales $\mu_R$ and $\mu_F$~\cite{Kalogeropoulos:2018cke}, and corrections to the treatment of initial-state radiation (ISR).
Most uncertainties affecting the signal prediction are treated as fully correlated among the analyses. 
The uncertainty in the modeling of the trigger efficiency is correlated between individual analyses that share primary high-level trigger paths. The uncertainty labeled as ``trigger timing'' relates to correction factors that were applied to account for a gradual shift in the timing of the trigger information from the ECAL toward early values, present during the 2016 and 2017 data-taking periods.
All analyses include the statistical uncertainty of the simulated signal
samples, which is taken as uncorrelated.

The dominant uncertainties in the background predictions are generally not correlated among analyses, as they tend to be either statistical in nature or inherent in the methods themselves. All systematic uncertainties associated with the normalization of SM backgrounds are taken to be uncorrelated, with the exception of the WZ normalization for the \multilep ~\cite{CMS:2021cox} and the ``$2/3\Pell$ soft''~\cite{CMS:2021edw} searches. 

\begin{table}[bp]
\topcaption{
  Sources of systematic uncertainties and the treatment of their correlations between analyses.  Exceptions to the notations in the last column are:  for the ``SM background normalization'' row, the $\PW\PZ$ normalization is correlated between the \multilep~\cite{CMS:2021cox} and ``$2/3\Pell$ soft''~\cite{CMS:2021edw} searches, and for the ``Lepton efficiency'' row the two contributing searches, ``$2/3\Pell$ soft''~\cite{CMS:2021edw} and ``$2\Pell$ nonresonant''~\cite{Sirunyan:2020eab}, are uncorrelated because they cover disjoint regions of the model parameter space.
}
\centering
\begin{scotch}{ll}
Source & Correlated?  \\
\multicolumn{2}{c}{General} \\
MC sample size                    & No  \\
SM background normalization     & No \\ 
Integrated luminosity              & Yes  \\
Trigger efficiency                & Partially \\
Pileup                            & Yes  \\
Trigger timing                & Yes \\
& \\
\multicolumn{2}{c}{Objects and signal modeling} \\
Lepton efficiency                 & Yes\\ 
Jet energy resolution             & Yes  \\
Jet energy scale        & Yes  \\
\PQb (mis)tagging efficiency      & Yes  \\
AK8 $\PQb\PQb$ tagging efficiency & Yes  \\
AK8 jet mass resolution           & Yes   \\
$\mu_R$ and $\mu_F$ & Partially   \\
ISR modeling             & Yes  \\
& \\
\multicolumn{2}{c}{Attributable to the CMS fast simulation} \\
\ptmiss modeling & Yes \\
\PQb (mis)tagging         & Yes  \\
AK8 $\PQb\PQb$ tagging    & Yes \\
AK8 $\PQb\PQb$ mass       & Yes \\
\end{scotch}
\label{table:systsCondensed}
\end{table}

\section{Results and interpretation}
\label{sec:interpretations}

The observed SR yields are incorporated into an ML fit, along with the predicted yields pertinent to a given model, for each point in the model parameter space considered.
The fit is implemented with the statistical framework described in Ref.~\cite{CMS-NOTE-2011-005}, which builds the likelihood as a product of Poisson probability density functions of the observed yields in all bins of all SRs.  The expectation in each bin is defined as the sum of background and signal contributions, with the signal contribution scaled by a strength parameter $\mu$ that is free in the fit.
The observed data yields in control regions are incorporated either by inclusion of the CRs in the likelihood or through parametrization using gamma functions.
Systematic uncertainties are taken as nuisance parameters in the fit, and are implemented using log-normal functions, whose widths reflect the size of the systematic uncertainty, or as Gaussian shape constraints based on the relevant distributions.

No significant deviations from the SM background are observed, consistent with the findings of the input searches.  Small tensions between the data and background seen in the original searches are generally reduced or are unchanged.
Cross section upper limits at 95\% confidence level (\CL) as a function of the SUSY particle masses are set using a modified frequentist approach, employing the \CLs criterion and an asymptotic formulation~\cite{Junk:1999kv,0954-3899-28-10-313,Cowan:2010js,CMS-NOTE-2011-005}.

The following paragraphs present the cross section limits as functions of one or two model parameters, together with exclusion limits given by the $\mu=1$ contours.  For visualization of the results, linear interpolation is employed  to account for the limited granularity of the available signal samples. 

\begin{figure*}[!ht]
  \centering
    \includegraphics[width=0.49\textwidth]{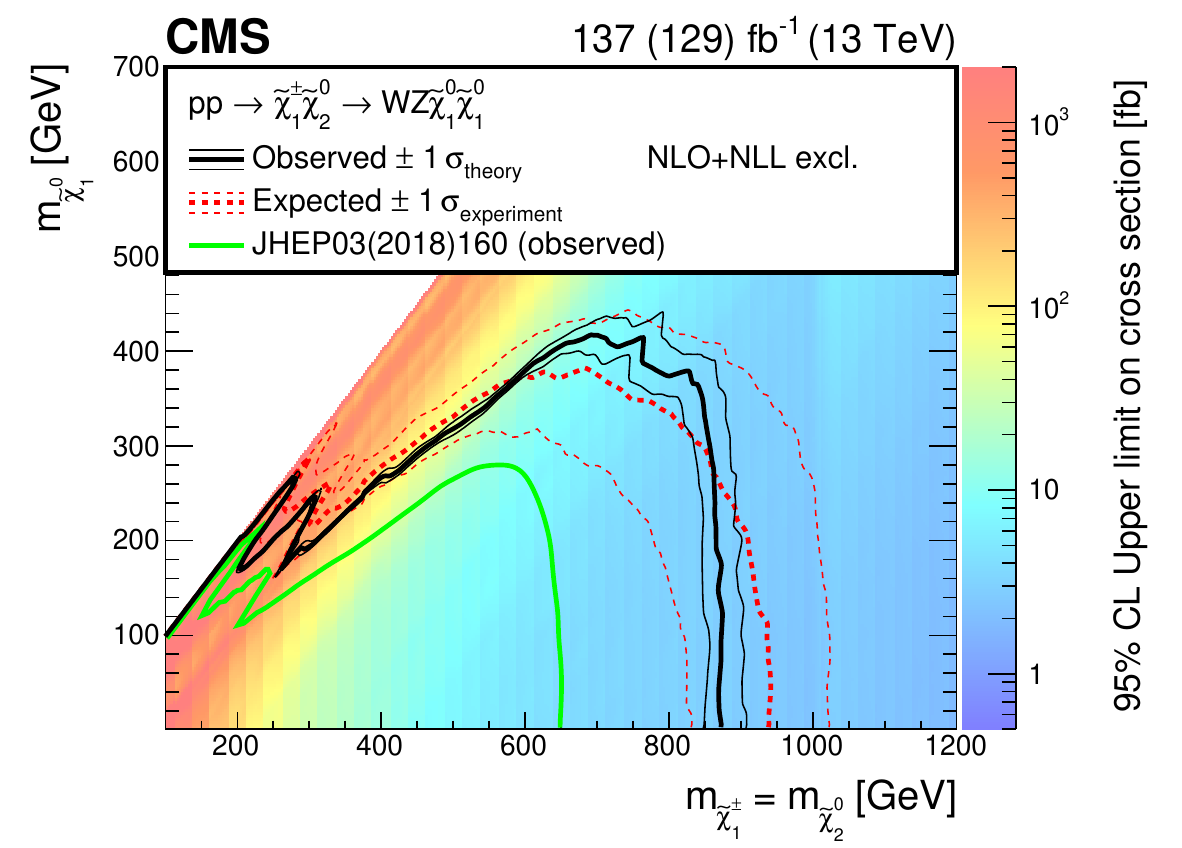}
    \includegraphics[width=0.49\textwidth]{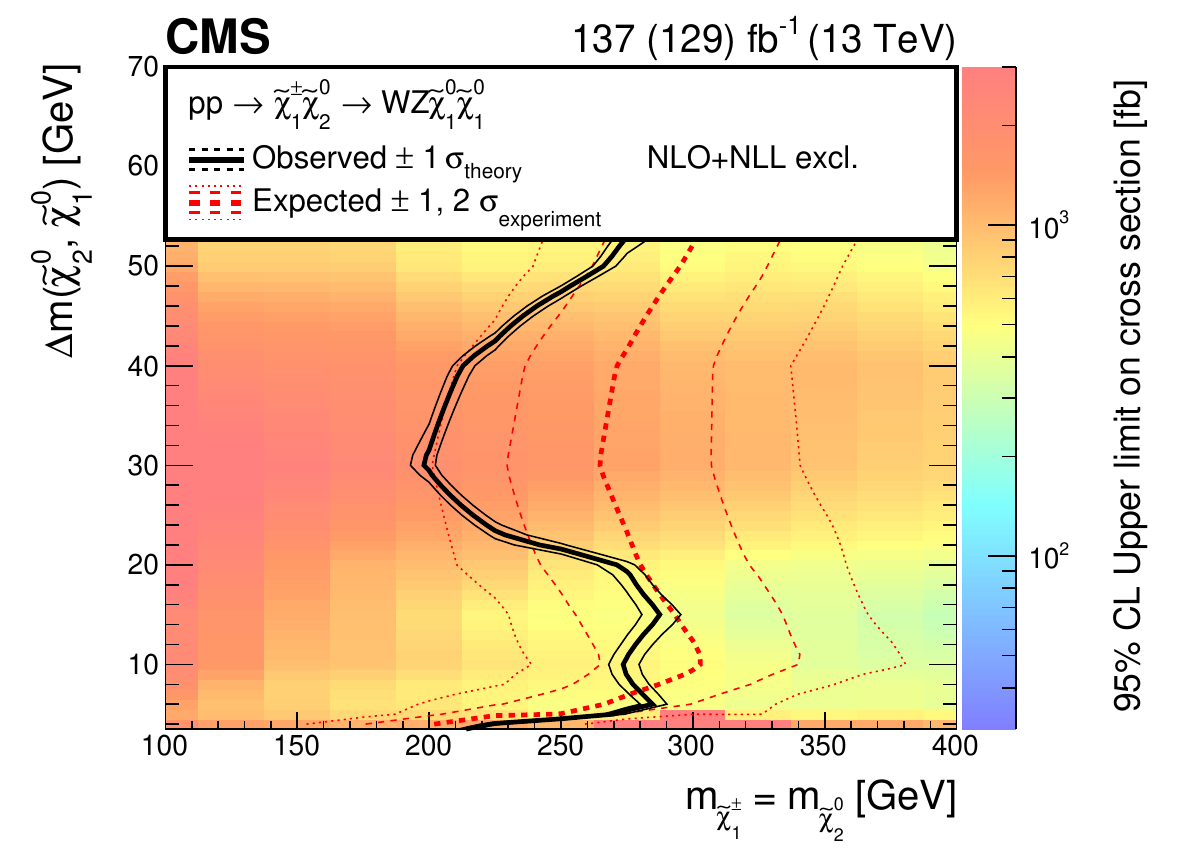}
    \includegraphics[width=0.49\textwidth]{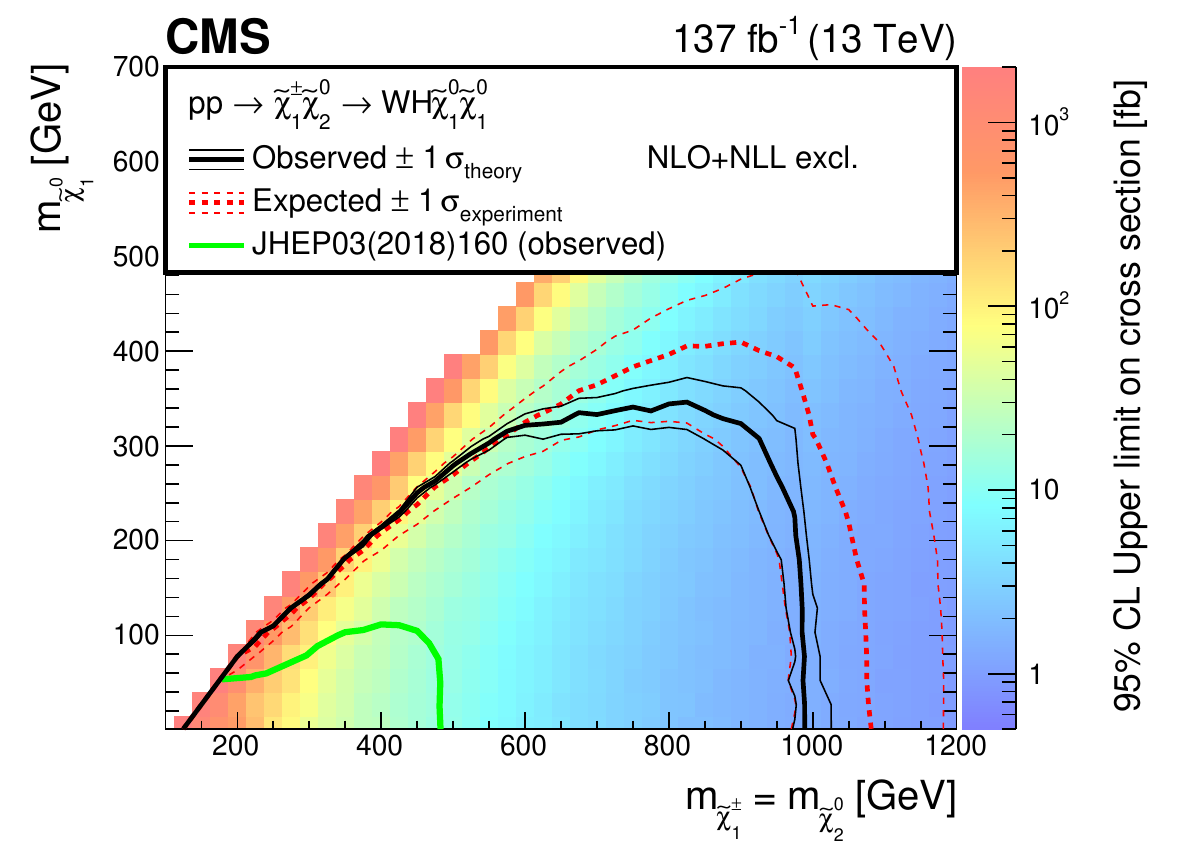}
    \includegraphics[width=0.49\textwidth]{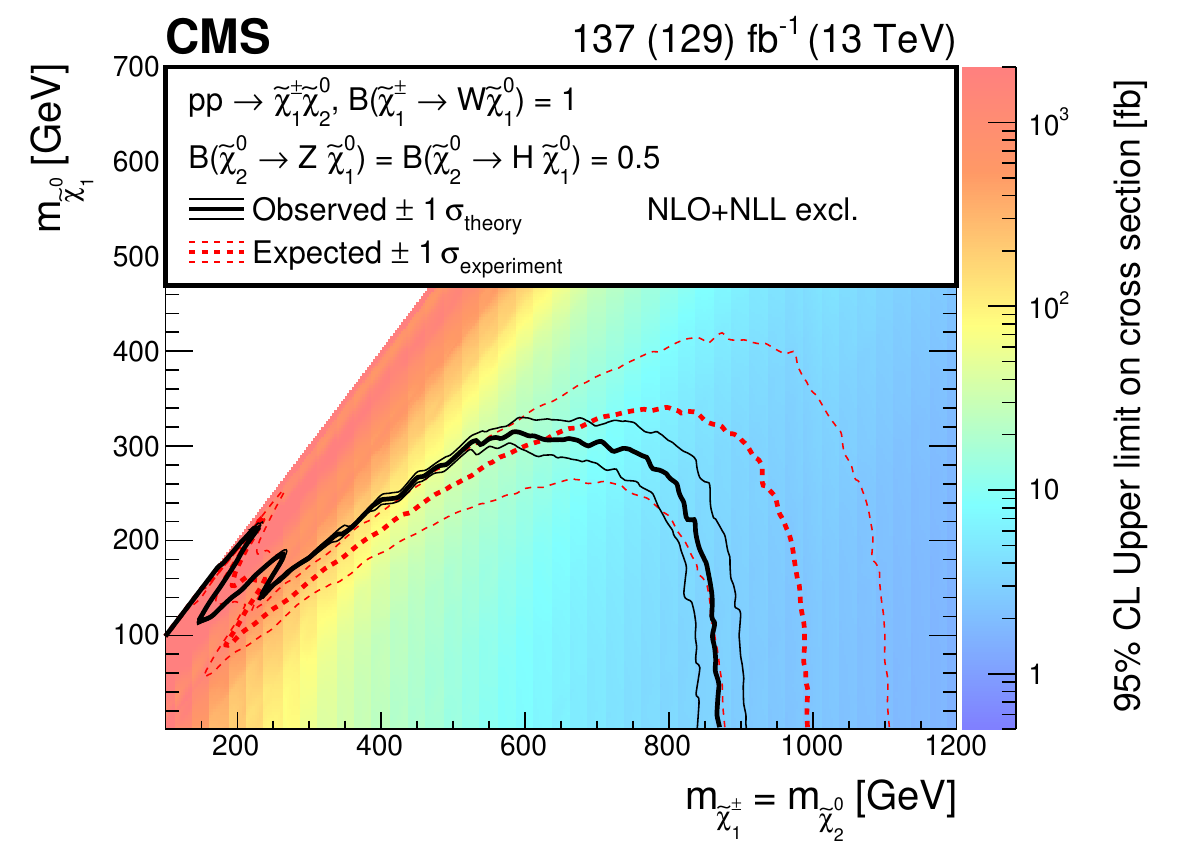}
  \caption{Wino-bino model: cross section limits with expected and observed exclusion boundaries in the model parameter space in the $\PW\PZ$ topology for the full parameter space (upper left) as well as the compressed region (upper right), the $\PW\PH$ topology (lower left), and the mixed topology with 50\% branching fraction to $\PW\PZ$ and $\PW\PH$ (lower right). All masses below the contours are excluded, except in the case of the upper-right plot where the area on the left of the contour is excluded. For some signal regions the analysis was based on a subset of the data, corresponding to the integrated luminosity of 129\fbinv.}
  \label{fig:TChiWX}
\end{figure*}

\begin{figure*}[!ht]
  \centering
    \includegraphics[width=0.49\textwidth]{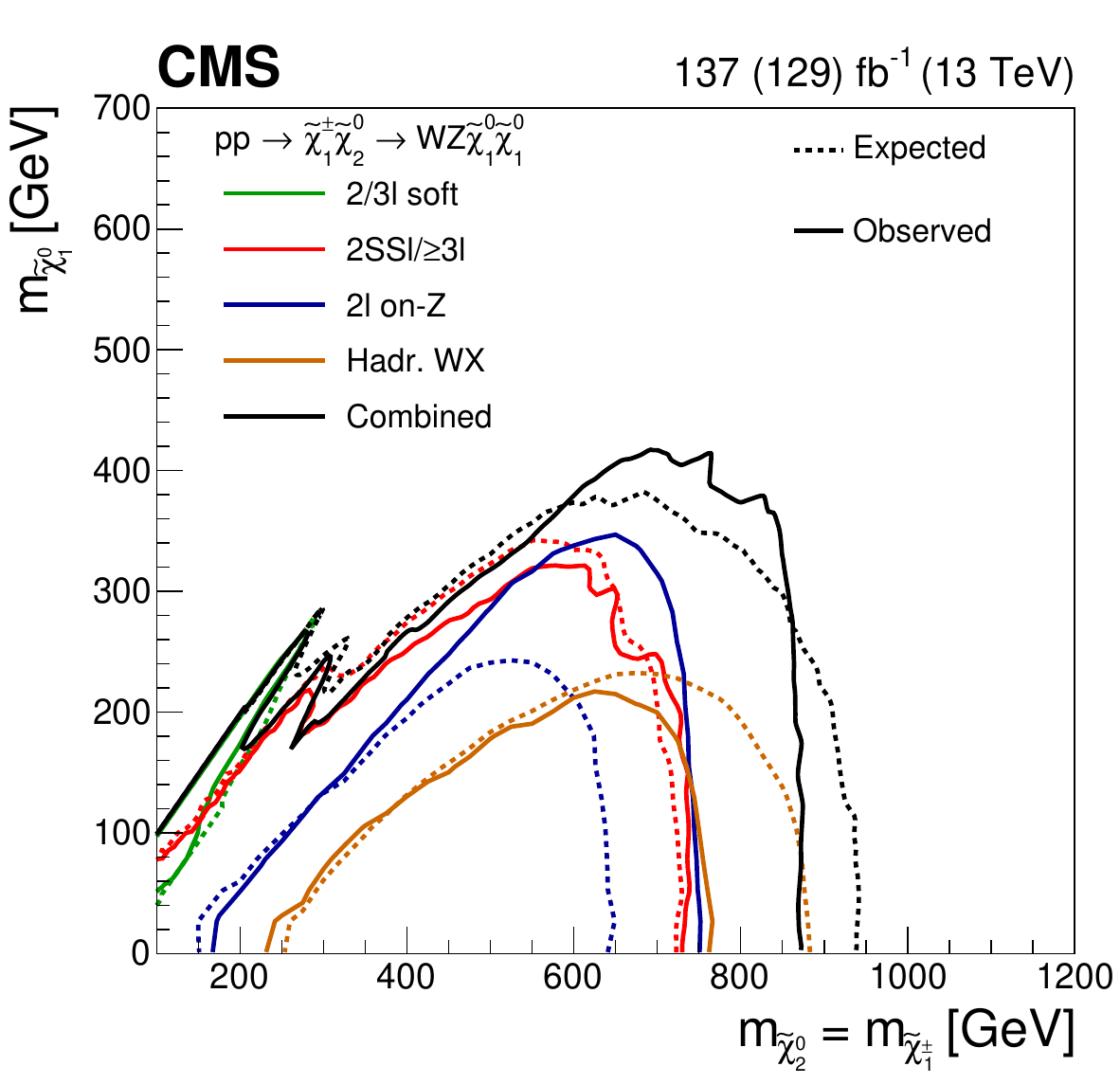}
    \includegraphics[width=0.49\textwidth]{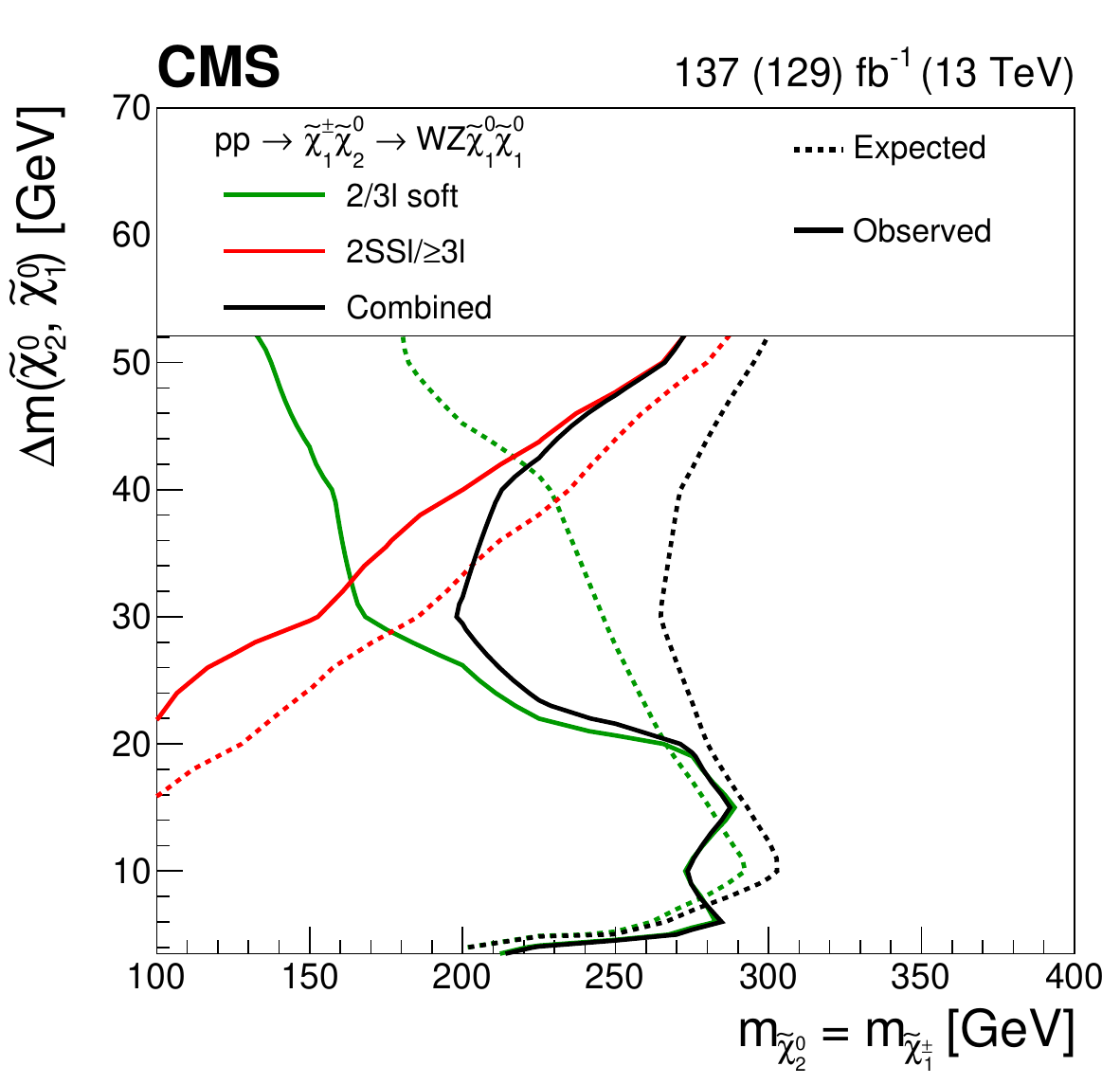}
    \includegraphics[width=0.49\textwidth]{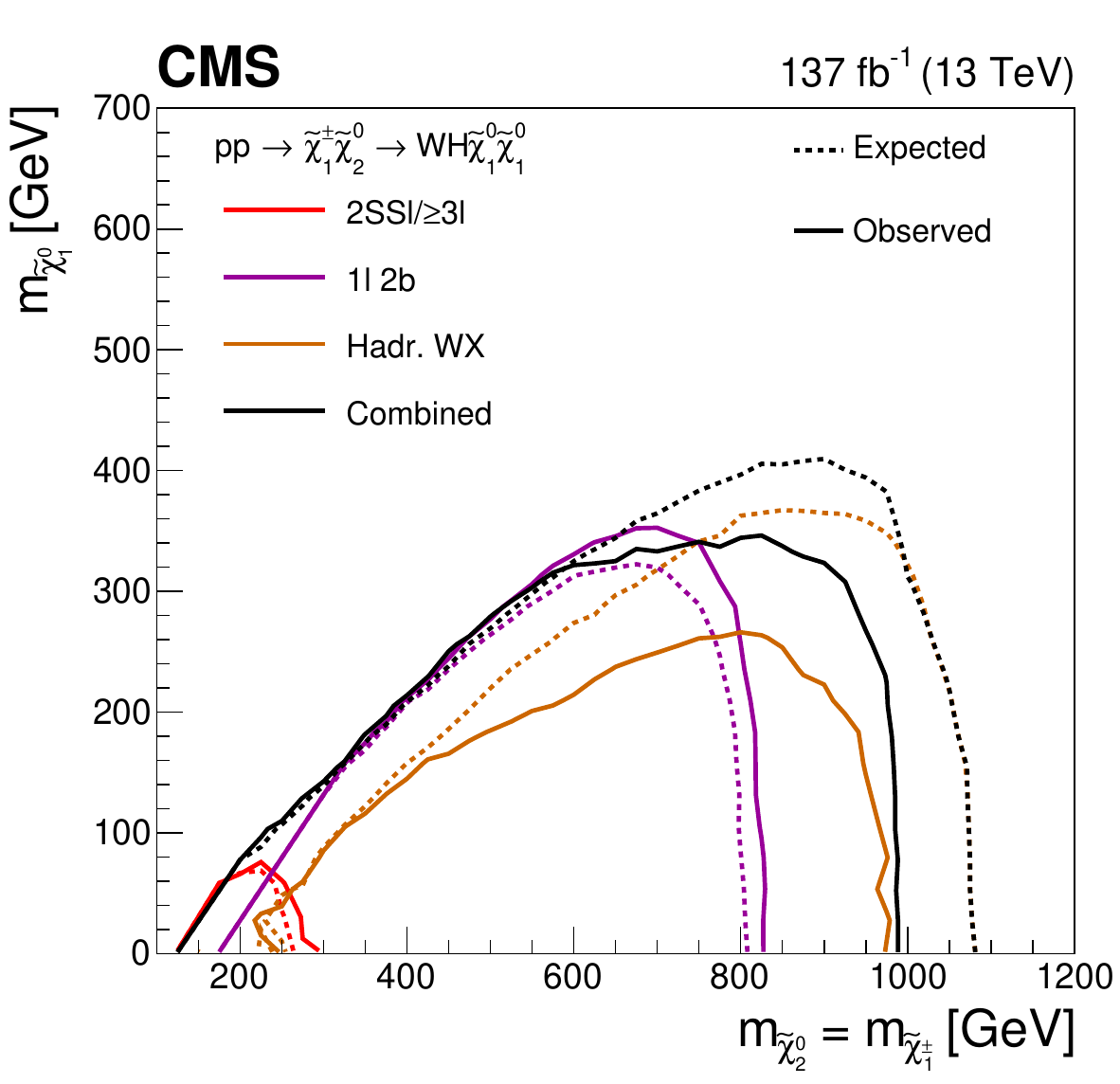}    
    \includegraphics[width=0.49\textwidth]{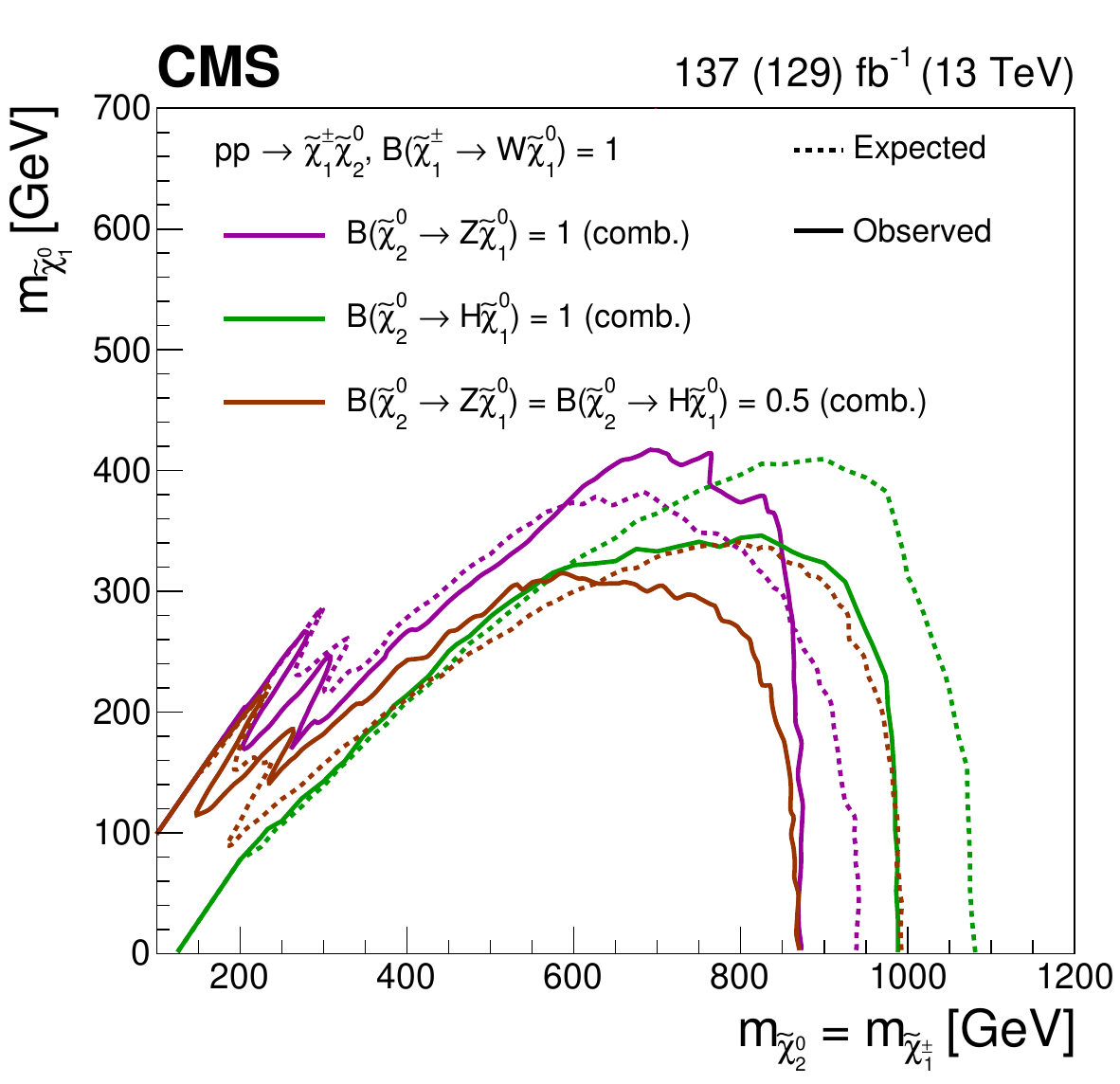}
  \caption{Wino-bino model: exclusion contours from the individual analyses targeting the $\PW\PZ$ topology for the full parameter space (upper left), the corresponding compressed region (upper right), and the $\PW\PH$ topology (lower left). The combined contours for these two topologies are also shown.  The combined contours for these and the mixed topology are overlaid in the lower-right plot. The decay modes assumed for each contour are given in the legends. All masses below the contours are excluded, except in the case of the upper-right plot where the area on the left of the contour is excluded. For some signal regions the analysis was based on a subset of the data, corresponding to the integrated luminosity of 129\fbinv.
}
\label{fig:TChiWX_individual}
\end{figure*}

The cross section upper limits and mass-limit contours for the wino-bino model with chargino-neutralino production (Fig.~\ref{fig:sigWX}) are shown in Fig.~\ref{fig:TChiWX}. 
The contributions of the input analyses to the limit curves in comparison with those from the combination appear in Fig.~\ref{fig:TChiWX_individual}, along with the combined mass plane limit contours for three values of the $\PH\ninoone / \PZ\ninoone$ branching ratio of the $\ninotwo$ decay. 

The distributions in Fig.~\ref{fig:TChiWX_individual} show that the excluded parameter space is expanded by the combination compared with that from the input analyses, rather substantially in the case of the $\PW\PZ$ topology in the uncompressed region (Fig.~\ref{fig:TChiWX_individual} upper left).  For that case the combined data exclude NLSP masses below 875\GeV for a light LSP, and LSP masses below 420 for a 700\GeV NLSP, gains of about 125 and 110\GeV, respectively.
In the compressed region, excesses observed in the combined data for the $\PW\PZ$ topology reduce the excluded space relative to the expected one by ${\approx}2$ standard deviations in the region of \dm around 30--40\GeV.  The upper-right plot in Fig.~\ref{fig:TChiWX_individual} shows that this falls in the crossover region where the exclusion power shifts between the two contributing analyses, ``2/3$\Pell$ soft" and \multilep. The less restrictive observed than expected exclusion in the region of large NLSP and LSP masses for the $\PW\PH$ and mixed topologies is driven by ${\approx}1$ standard deviation excesses in several \ptmiss bins of the $\PW\PH$ SR in the ``Hadr.~WX'' analysis, as demonstrated by the curves in Fig.~\ref{fig:TChiWX_individual}, lower left.

\begin{figure}[!ht]
  \centering
    \includegraphics[width=0.49\textwidth]{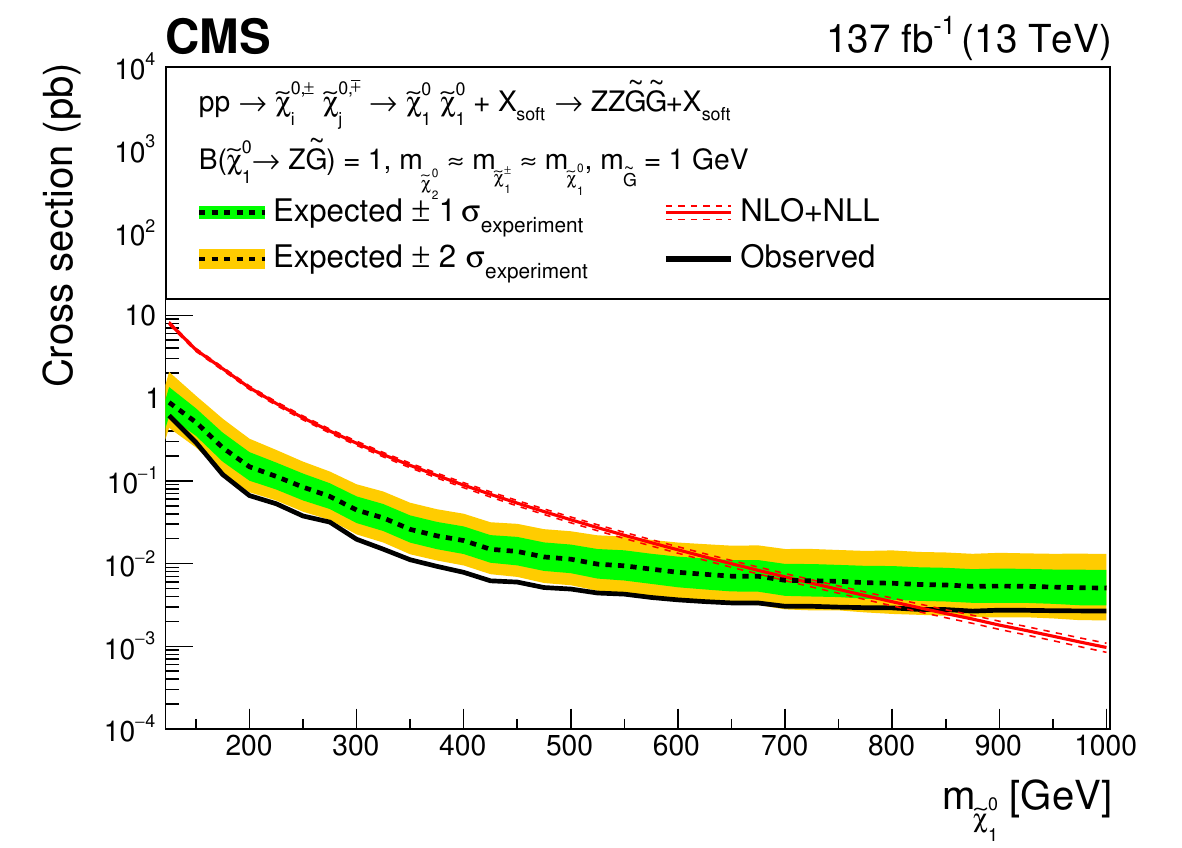}\\
    \includegraphics[width=0.49\textwidth]{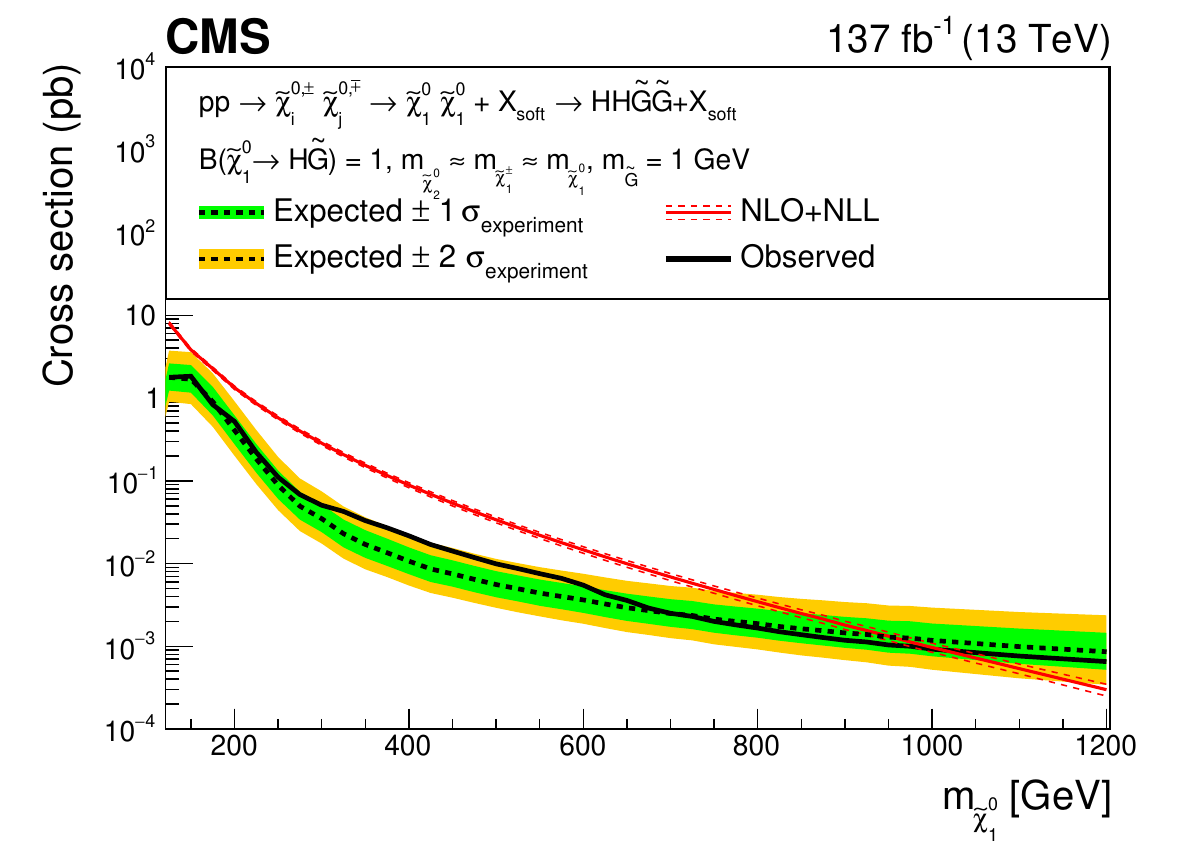}\\
    \includegraphics[width=0.49\textwidth]{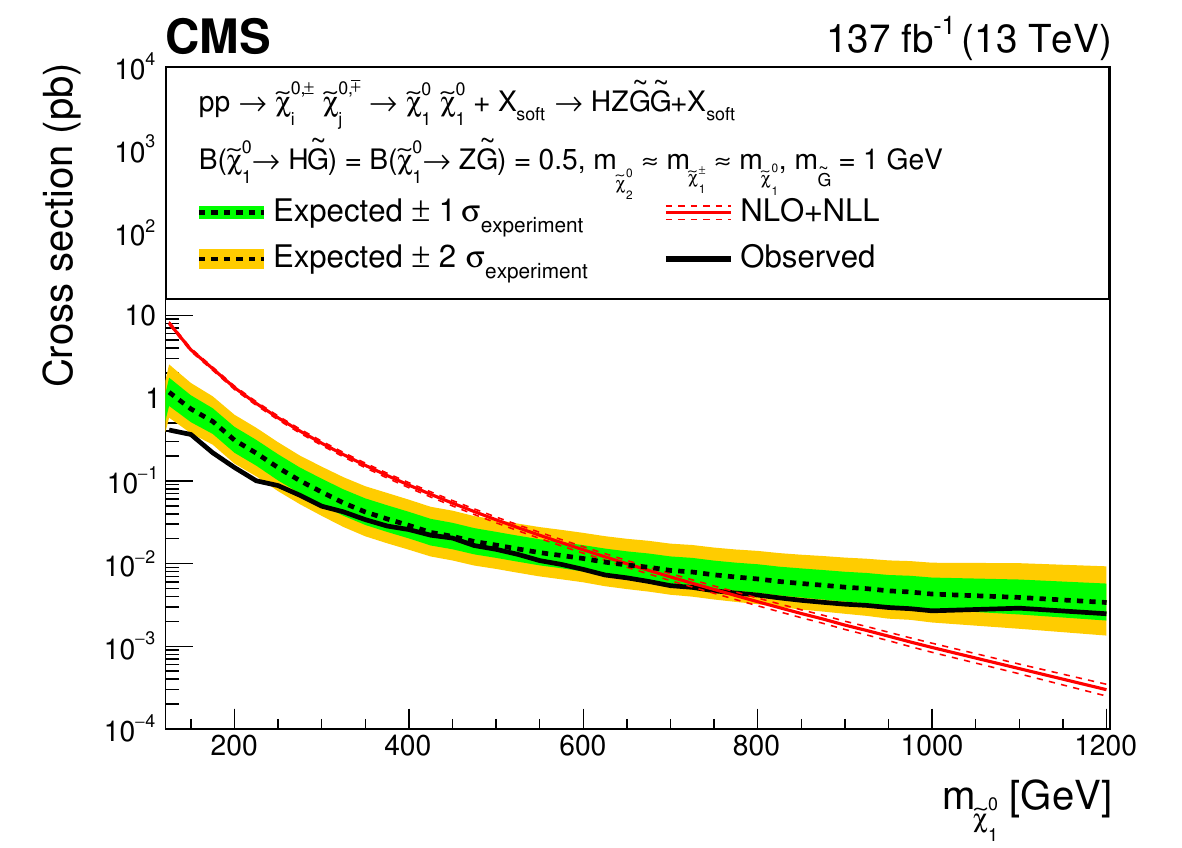}
  \caption{GMSB model: expected and observed exclusion limits for the $\PZ\PZ$ topology (upper), the $\PH\PH$ topology (middle), and the mixed topology with 50\% branching fraction to \PH and \PZ (lower). All masses on the left of the crossing between the exclusion limits and theory prediction are excluded.}
  \label{fig:TChiNN}
\end{figure}

\begin{figure}[!ht]
  \centering
    \includegraphics[width=0.49\textwidth]{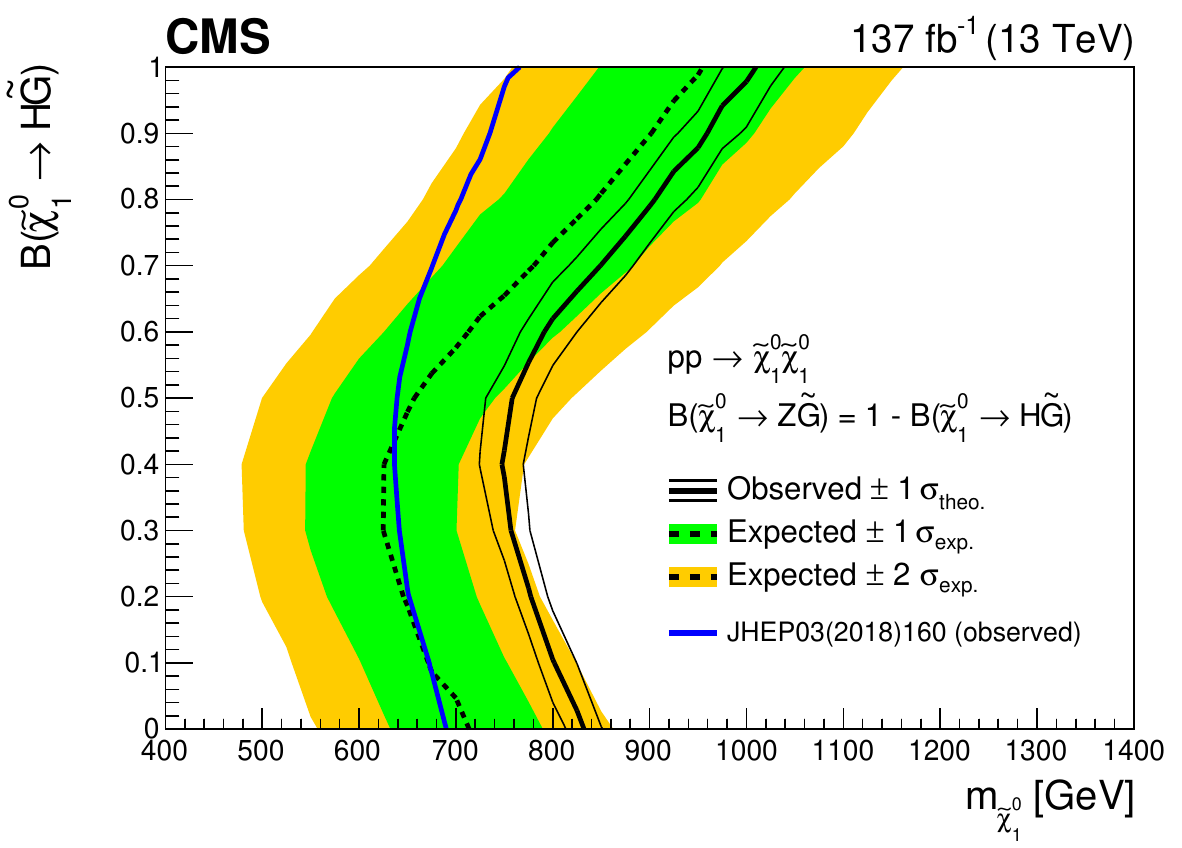}
     \includegraphics[width=0.49\textwidth]{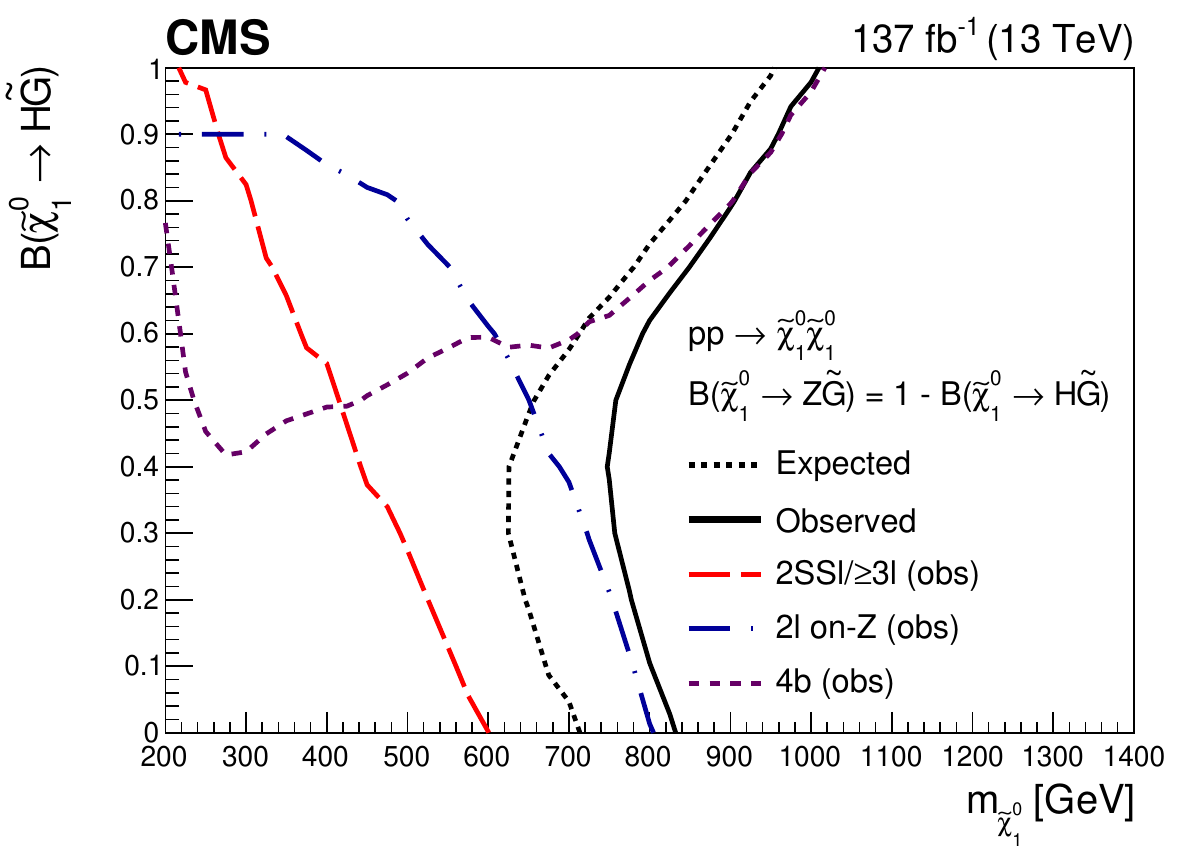}
  \caption{GMSB model: NLSP mass exclusion limit as a function of the branching fraction to the \PH boson. \cmsLEFT: expected and observed limits for the combination of the searches, shown together with the observed limits of the combination~\cite{Sirunyan:2018ubx} based on the 2016 CMS data.
    \cmsRIGHT: expected and observed exclusion limits for the combination in comparison with those of the input searches. All masses on the left of the contours are excluded.}
  \label{fig:TChiNN_BR}
\end{figure}

For the GMSB model (Fig.~\ref{fig:sigNeutral}) with production of degenerate higgsinos decaying to the goldstino \gravitino (neutralino-neutralino production),
the cross section upper limits and exclusion curves are shown in Fig.~\ref{fig:TChiNN}, for the topologies $\PZ\PZ$, $\PH\PH$, and the mixture of these with $\PH\PZ$ assuming $\mathcal{B}(\ninoone\to\PZ\gravitino) = \mathcal{B}(\ninoone\to\PH\gravitino) = 50\%$.  For the mixed topology, we also provide, in Fig.~\ref{fig:TChiNN_BR}, exclusion limits spanning the full range of branching fractions into an \PH and \PZ boson. 
Mass values for \ninoone less than 840\GeV are excluded for $\PZ\PZ$, ${<}\,1025\GeV$ for the $\PH\PH$ topology, and ${<}\,760\GeV$ for the 50\% mixed topology.  The impact of the combination is substantial for the mixed topology, increasing the excluded region by about 100\GeV. The $\PZ\PZ$ and mixed topologies show better than expected exclusion, whereas the $\PH\PH$ topology shows the impact of a one-bin excess from the ``4b" analysis, a 1--2 standard deviation effect for $300<\mninoone<600\GeV$.

\begin{figure}[!ht]
  \centering
  \includegraphics[width=0.49\textwidth]{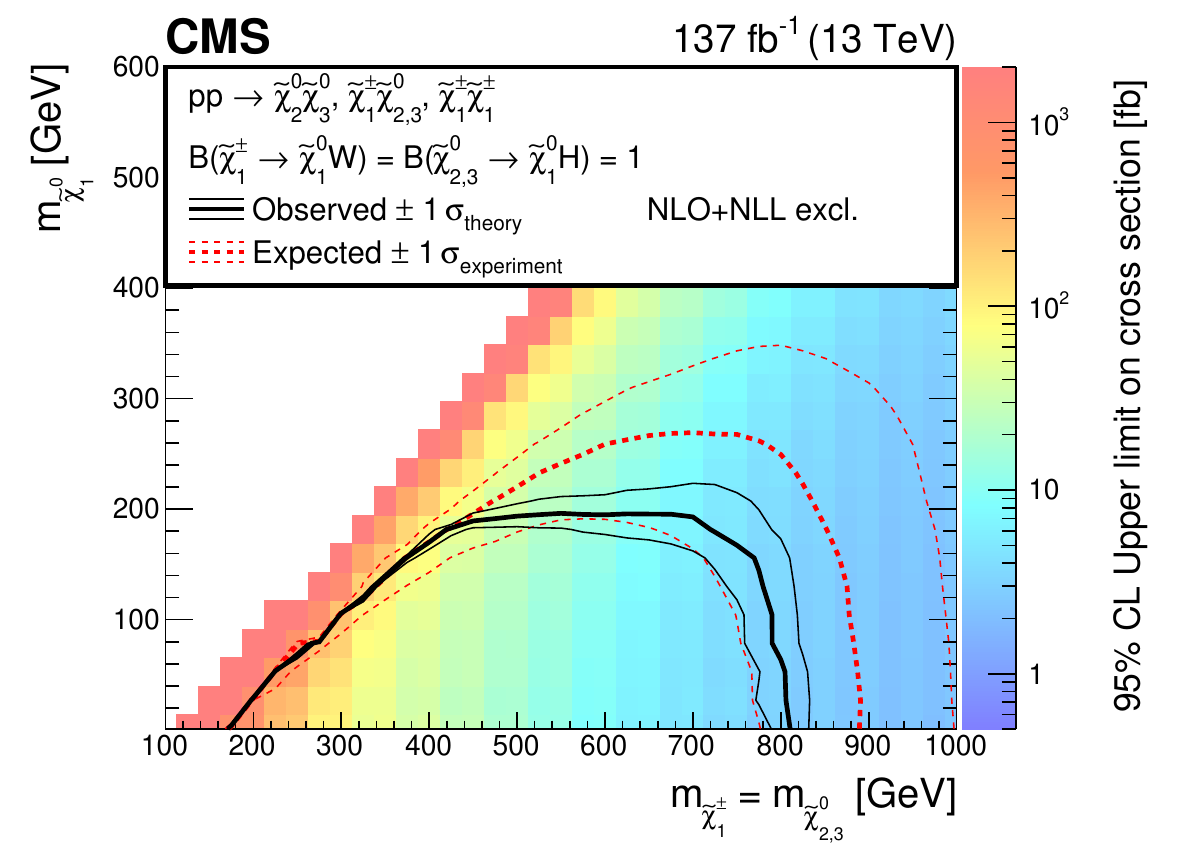}
  \caption{Higgsino-bino model: cross section upper limit in the mass plane of the model, and the expected and observed exclusion limits.  The model assumes mass-degenerate higgsino-like \ninotwo, \ninothree, and \chinoone that decay to a bino-like \ninoone and an SM boson. All masses below the contours are excluded.}
  \label{fig:higgsino-bino}
\end{figure}

The limits for the higgsino-bino interpretation (Fig.~\ref{fig:higgsino_bino}), which assume mass-degenerate higgsino-like \ninotwo, \ninothree, and \chinoone that decay to a bino-like \ninoone and an SM boson, are shown in Fig.~\ref{fig:higgsino-bino}.  Values of $\mchinoone=\mninotwo$ between 225 and 800\GeV are excluded for \ninoone masses below 50\GeV.  As noted, this interpretation was not addressed in the previous analyses.

\begin{figure}[!ht]
  \centering
  \includegraphics[width=0.49\textwidth]{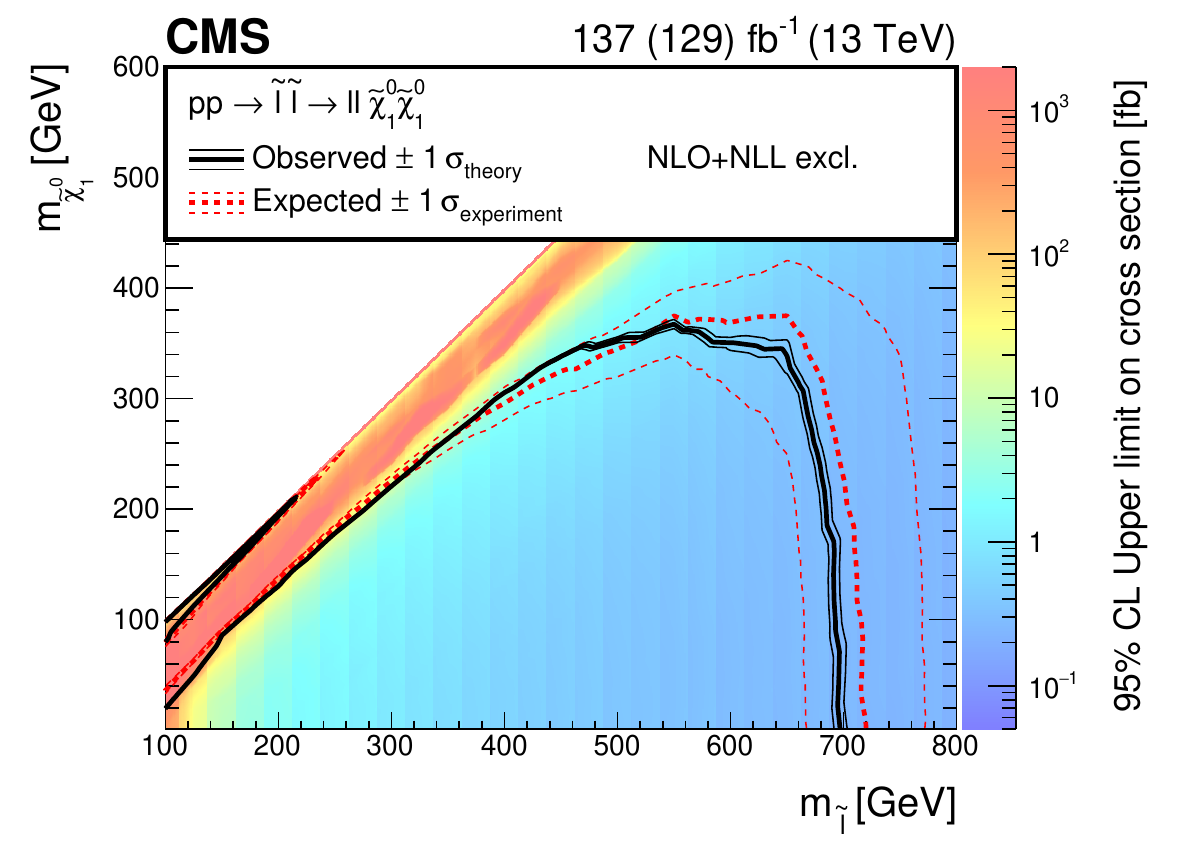}
  \includegraphics[width=0.49\textwidth]{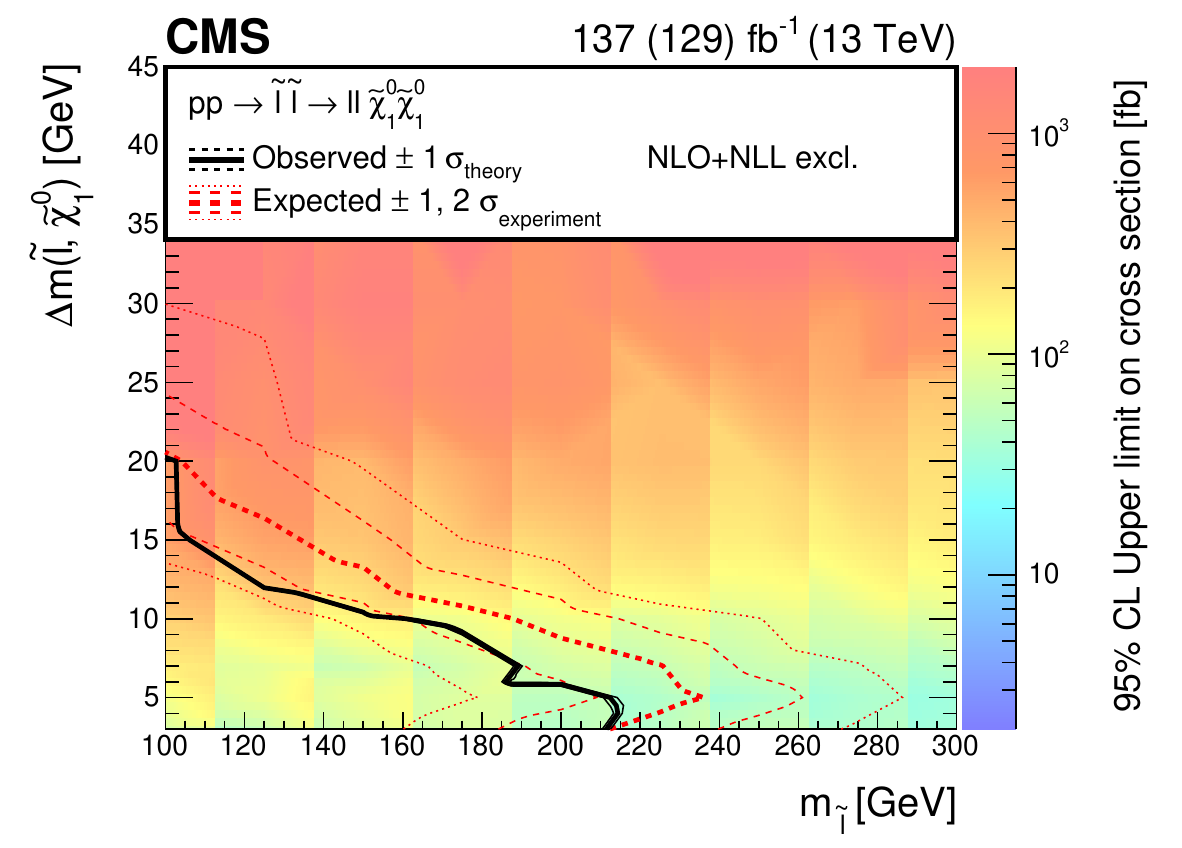}
  \caption{Slepton-neutralino model: mass plane cross section upper limit with observed and expected exclusion limits.  \cmsLEFT:  the full mass plane from the combination.
\cmsRIGHT:  the compressed region, obtained by the ``$2/3\Pell$ soft'' search. For some signal regions the analysis was based on a subset of the data, corresponding to the integrated luminosity of 129\fbinv. All masses below the contours are excluded, except in the case of the right figure where the area on the left of the contour is excluded.}
  \label{fig:sleptons}
\end{figure}

The limits for slepton pair production under the slepton-neutralino model (Fig.~\ref{fig:sigSlepton}) are shown in Fig.~\ref{fig:sleptons}.
The left-hand plot shows the combined limit over the $(m_{\slep},m_{\ninoone})$ plane, while the right-hand plot gives an expanded view of the compressed region.
In this region the ``$2/3\Pell$ soft'' search is able to exclude values of slepton masses up to 215\GeV at $\dm=5\GeV$, complementing the results in the uncompressed region of the model parameter space where the observed exclusion reaches the slepton mass of about 700\GeV.

\section{Summary}
\label{sec:summary}

A number of previously reported searches for supersymmetry (SUSY) in different final states from proton-proton
collisions at $\sqrt{s}=13$ TeV have been reoptimized and combined.  The data were recorded with the CMS detector at the LHC and
correspond to an integrated luminosity of up to 137 \fbinv.
These data are used to test the predictions of
a variety of simplified SUSY models that involve the
electroweak production of the superpartners of electroweak gauge or Higgs bosons.
No significant deviation from the standard model expectation has been observed,
and limits are set on the production of
wino-like chargino-neutralino pairs, higgsino-like neutralino pair production
in a gauge-mediated SUSY breaking inspired scenario, a higgsino-bino interpretation, and slepton pair production.

In the case of wino-like chargino-neutralino production, for a \ninoone, the lightest SUSY particle in this model, with mass $\mchinoone<50\GeV$, the combined result gives an observed (expected) limit on \mchinoone  of about 875 (950)\GeV for the $\PW\PZ$ topology, 990 (1075)\GeV for the $\PW\PH$ topology, and 875 (1000)\GeV for a mixed topology, extending the previous CMS combination~\cite{Sirunyan:2018ubx} (based on a 2016 data set corresponding to $36\fbinv$),
by 225, 510, and 340\GeV, respectively, for the three topologies.

For higgsino-like neutralino pair production, the mass exclusion limit is a function of the branching ratio between the \PH and \PZ channels; 
the expected limit ranges between
about 620 and 950\GeV, the smaller value occurring for $\mathcal{B}(\ninoone \to \PH\gravitino) \approx 0.4$. 
For this value of the branching fraction, the observed limit results in the exclusion of masses below 750\GeV, and extends the previous result~\cite{Sirunyan:2018ubx} by 100\GeV. 
The observed limit reaches nearly 1025\GeV at $\mathcal{B}(\ninoone \to \PH\gravitino) = 1$, to be compared with 750\GeV reported in Ref.~\cite{Sirunyan:2018ubx}.

A higgsino-bino model that assumes mass degenerate higgsino-like \ninotwo, \ninothree, and \chinoone decaying to a bino-like \ninoone and a standard model boson is excluded for $\mchinoone=\mninotwo$ between 225 and 800\GeV for $\mninoone<50\GeV$.

For direct production of the superpartners of electrons and muons (sleptons),
this combined search yields an observed (expected) exclusion in the slepton mass of about 130--700 (110--720)\GeV, for $\mninoone<50\GeV$.  In the compressed-spectrum region, a dedicated search first reported here excludes slepton masses above 215\GeV for a 5\GeV difference between the slepton and \ninoone masses.

In general for the models considered in this combination, wino-like chargino masses are excluded up to 990\GeV, while higgsino-like neutralino masses are excluded up to 1025\GeV. 
The improvement is between 100--510\GeV with respect to the previous exclusion limits~\cite{Sirunyan:2018ubx}, 
whereas the excluded model parameter space is expanded by as much as 125\GeV, depending on the model, with respect to the best of the component searches. The compressed parameter space of the slepton production model is explored here for the first time by CMS.

\begin{acknowledgments}
  We congratulate our colleagues in the CERN accelerator departments for the excellent performance of the LHC and thank the technical and administrative staffs at CERN and at other CMS institutes for their contributions to the success of the CMS effort. In addition, we gratefully acknowledge the computing centers and personnel of the Worldwide LHC Computing Grid and other centers for delivering so effectively the computing infrastructure essential to our analyses. Finally, we acknowledge the enduring support for the construction and operation of the LHC, the CMS detector, and the supporting computing infrastructure provided by the following funding agencies: SC (Armenia), BMBWF and FWF (Austria); FNRS and FWO (Belgium); CNPq, CAPES, FAPERJ, FAPERGS, and FAPESP (Brazil); MES and BNSF (Bulgaria); CERN; CAS, MoST, and NSFC (China); MINCIENCIAS (Colombia); MSES and CSF (Croatia); RIF (Cyprus); SENESCYT (Ecuador); ERC PRG, RVTT3 and TK202 (Estonia); Academy of Finland, MEC, and HIP (Finland); CEA and CNRS/IN2P3 (France); SRNSF (Georgia); BMBF, DFG, and HGF (Germany); GSRI (Greece); NKFIH (Hungary); DAE and DST (India); IPM (Iran); SFI (Ireland); INFN (Italy); MSIP and NRF (Republic of Korea); MES (Latvia); LAS (Lithuania); MOE and UM (Malaysia); BUAP, CINVESTAV, CONACYT, LNS, SEP, and UASLP-FAI (Mexico); MOS (Montenegro); MBIE (New Zealand); PAEC (Pakistan); MES and NSC (Poland); FCT (Portugal); MESTD (Serbia); MCIN/AEI and PCTI (Spain); MOSTR (Sri Lanka); Swiss Funding Agencies (Switzerland); MST (Taipei); MHESI and NSTDA (Thailand); TUBITAK and TENMAK (Turkey); NASU (Ukraine); STFC (United Kingdom); DOE and NSF (USA).

\hyphenation{Rachada-pisek} Individuals have received support from the Marie-Curie program and the European Research Council and Horizon 2020 Grant, contract Nos.\ 675440, 724704, 752730, 758316, 765710, 824093, and COST Action CA16108 (European Union); the Leventis Foundation; the Alfred P.\ Sloan Foundation; the Alexander von Humboldt Foundation; the Science Committee, project no. 22rl-037 (Armenia); the Belgian Federal Science Policy Office; the Fonds pour la Formation \`a la Recherche dans l'Industrie et dans l'Agriculture (FRIA-Belgium); the Agentschap voor Innovatie door Wetenschap en Technologie (IWT-Belgium); the F.R.S.-FNRS and FWO (Belgium) under the ``Excellence of Science -- EOS" -- be.h project n.\ 30820817; the Beijing Municipal Science \& Technology Commission, No. Z191100007219010 and Fundamental Research Funds for the Central Universities (China); the Ministry of Education, Youth and Sports (MEYS) of the Czech Republic; the Shota Rustaveli National Science Foundation, grant FR-22-985 (Georgia); the Deutsche Forschungsgemeinschaft (DFG), under Germany's Excellence Strategy -- EXC 2121 ``Quantum Universe" -- 390833306, and under project number 400140256 - GRK2497; the Hellenic Foundation for Research and Innovation (HFRI), Project Number 2288 (Greece); the Hungarian Academy of Sciences, the New National Excellence Program - \'UNKP, the NKFIH research grants K 124845, K 124850, K 128713, K 128786, K 129058, K 131991, K 133046, K 138136, K 143460, K 143477, 2020-2.2.1-ED-2021-00181, and TKP2021-NKTA-64 (Hungary); the Council of Science and Industrial Research, India; ICSC -- National Research Center for High Performance Computing, Big Data and Quantum Computing, funded by the EU NexGeneration program (Italy); the Latvian Council of Science; the Ministry of Education and Science, project no. 2022/WK/14, and the National Science Center, contracts Opus 2021/41/B/ST2/01369 and 2021/43/B/ST2/01552 (Poland); the Funda\c{c}\~ao para a Ci\^encia e a Tecnologia, grant CEECIND/01334/2018 (Portugal); the National Priorities Research Program by Qatar National Research Fund; MCIN/AEI/10.13039/501100011033, ERDF ``a way of making Europe", and the Programa Estatal de Fomento de la Investigaci{\'o}n Cient{\'i}fica y T{\'e}cnica de Excelencia Mar\'{\i}a de Maeztu, grant MDM-2017-0765 and Programa Severo Ochoa del Principado de Asturias (Spain); the Chulalongkorn Academic into Its 2nd Century Project Advancement Project, and the National Science, Research and Innovation Fund via the Program Management Unit for Human Resources \& Institutional Development, Research and Innovation, grant B37G660013 (Thailand); the Kavli Foundation; the Nvidia Corporation; the SuperMicro Corporation; the Welch Foundation, contract C-1845; and the Weston Havens Foundation (USA).
\end{acknowledgments}
\bibliography{auto_generated}
\cleardoublepage \appendix\section{The CMS Collaboration \label{app:collab}}\begin{sloppypar}\hyphenpenalty=5000\widowpenalty=500\clubpenalty=5000\input{SUS-21-008-public-authorlist.tex}\end{sloppypar}
%%% END EDITABLE REGION %%%
% skeleton_end
\end{document}

%% file: SUS-21-008-public-authorlist.tex
\cmsinstitute{Yerevan Physics Institute, Yerevan, Armenia}
{\tolerance=6000
A.~Hayrapetyan, A.~Tumasyan\cmsAuthorMark{1}\cmsorcid{0009-0000-0684-6742}
\par}
\cmsinstitute{Institut f\"{u}r Hochenergiephysik, Vienna, Austria}
{\tolerance=6000
W.~Adam\cmsorcid{0000-0001-9099-4341}, J.W.~Andrejkovic, T.~Bergauer\cmsorcid{0000-0002-5786-0293}, S.~Chatterjee\cmsorcid{0000-0003-2660-0349}, K.~Damanakis\cmsorcid{0000-0001-5389-2872}, M.~Dragicevic\cmsorcid{0000-0003-1967-6783}, A.~Escalante~Del~Valle\cmsorcid{0000-0002-9702-6359}, P.S.~Hussain\cmsorcid{0000-0002-4825-5278}, M.~Jeitler\cmsAuthorMark{2}\cmsorcid{0000-0002-5141-9560}, N.~Krammer\cmsorcid{0000-0002-0548-0985}, D.~Liko\cmsorcid{0000-0002-3380-473X}, I.~Mikulec\cmsorcid{0000-0003-0385-2746}, J.~Schieck\cmsAuthorMark{2}\cmsorcid{0000-0002-1058-8093}, R.~Sch\"{o}fbeck\cmsorcid{0000-0002-2332-8784}, D.~Schwarz\cmsorcid{0000-0002-3821-7331}, M.~Sonawane\cmsorcid{0000-0003-0510-7010}, S.~Templ\cmsorcid{0000-0003-3137-5692}, W.~Waltenberger\cmsorcid{0000-0002-6215-7228}, C.-E.~Wulz\cmsAuthorMark{2}\cmsorcid{0000-0001-9226-5812}
% ...
\par}
\vskip\cmsinstskip
\dag:~Deceased\\
$^{1}$Also at Yerevan State University, Yerevan, Armenia\\
$^{2}$Also at TU Wien, Vienna, Austria\\
% ...
